\documentclass[prb,twocolumn]{revtex4}%

\usepackage{graphicx}
\usepackage{amsmath}
\usepackage{ulem}
\usepackage{color}

\newcommand{\be}{\begin{equation}}
\newcommand{\ee}{\end{equation}}
\newcommand{\bea}{\begin{eqnarray*}}
\newcommand{\eea}{\end{eqnarray*}}
\newcommand{\bean}{\begin{eqnarray}}
\newcommand{\eean}{\end{eqnarray}}

\begin{document}

\draft
\title{\bf End-State-Controlled Quantum Transport in Armchair Graphene Nanoribbon Artificial Quantum Materials}

\author{David M T Kuo}

\address{Department of Electrical Engineering and Department of Physics, National Central
University, Chungli, 32001 Taiwan}

\date{\today}

\begin{abstract}
Artificial quantum materials based on atomically precise graphene
nanostructures provide an ideal platform for exploring quantum
phenomena arising from localized electronic states. Here, we
develop a real-space theoretical framework to elucidate the
microscopic origin of interface states in graphene architectures
composed of $n$-triangulenes and armchair graphene nanoribbons
(AGNRs). By continuously tuning the coupling between graphene
building blocks, we reveal the evolution of triangulene
zero-energy modes and AGNR end states into compact localized node
orbitals at three-arm junctions. For the chiral bipartite
junctions considered here, the number and sublattice character of
these node orbitals follow a general counting relation,
$N_{node,\delta}=|N_{es,t,A(B)}-N_{tri,0,B(A)}|$, where $N_{es,t}$
is the total number of AGNR end states contributed by the three
AGNR arms and $N_{tri,0}$ is the number of triangulene zero-energy
modes. The resulting node-orbital chirality is determined by the
dominant constituent: when $N_{es,t}>N_{tri,0}$, they inherit the
sublattice chirality of the AGNR end states ($\delta=A(B)$),
whereas for $N_{tri,0}>N_{es,t}$ they inherit that of the
triangulene zero-energy modes ($\delta=B(A)$). Using
experimentally synthesized triangulene nanographenes as
representative examples, we further identify their low-energy
zero-mode structure and investigate its manifestation in tunneling
transport within an extended Anderson framework. Finally, we
demonstrate that these node orbitals can serve as elementary
building blocks for constructing artificial graphene nanoribbons
with highly tunable flat subbands near the Fermi energy. The
resulting compact localized states exhibit controllable degeneracy
and strongly anisotropic quantum transport. Our results establish
a general strategy for end-state orbital engineering in
graphene-based artificial quantum materials, in which the symmetry
and connectivity of molecular zero modes provide direct control
over localized states, flat-subband degeneracy and quantum
transport.
\end{abstract}

\maketitle

\section{Introduction}
Graphene possesses exceptionally high carrier mobility owing to
its gapless Dirac-cone electronic structure
[\onlinecite{Novoselovks}]. Controlling its electronic properties
through nanoscale structural engineering is therefore important
for applications in nanoelectronics, optoelectronics, and
thermoelectric devices. Among graphene nanostructures, armchair
graphene nanoribbons (AGNRs)
[\onlinecite{Cai}-\onlinecite{Nestor}], AGNR heterostructures
[\onlinecite{Groning}-\onlinecite{SunQ}], and Janus graphene
nanoribbon segments [\onlinecite{SongST}] exhibit tunable
electronic band gaps determined by their atomic structures. In
particular, AGNRs and their heterostructures provide a versatile
platform for engineering localized quantum states in atomically
precise nanostructures. Zero-energy in-gap states localized at
ribbon ends or heterostructure interfaces have been observed
experimentally through scanning tunneling microscopy and
spectroscopy [\onlinecite{Groning}--\onlinecite{DJRizzo}]. These
states, commonly associated with the topological properties of
AGNR-based structures, can be separated from bulk bands by energy
gaps exceeding one electron volt
[\onlinecite{DRizzo},\onlinecite{DJRizzo}], making them promising
candidates for realizing topological quantum dots and quantum
electronic devices
[\onlinecite{LeobandungE}--\onlinecite{PerrinML}]. Recent advances
in bottom-up synthesis have further enabled the fabrication of
triangulene-based graphene nanostructures with complex geometries,
opening new opportunities for engineering localized states beyond
conventional one-dimensional
nanoribbons[\onlinecite{HenriquesJ}--\onlinecite{YanY}].

An important issue in understanding these systems is how localized
states evolve when different graphene components are connected. In
our previous study of $N$-AGNR/$(N-2)$-AGNR/$N$-AGNR
heterostructures [\onlinecite{DavidKMT}], we investigated the
hybridization between different end states at the junctions and
its influence on their in-gap character. This work provides a
physical basis for examining how localized states respond to the
coupling between neighboring graphene components. A related
question arises in triangulene-based
Y-junctions[\onlinecite{TamakiG}], where the zero-energy modes of
the triangulene core
coexist[\onlinecite{GuYW}--\onlinecite{MadailL}] and hybridize
with the end states of the three attached AGNR segments.
Understanding this hybridization is essential for identifying the
localized orbitals formed at the junction and for determining how
these orbitals can be used to construct artificial graphene
nanoribbons with tunable electronic properties.

Previous theoretical studies have employed symmetry considerations,
including the threefold rotational symmetry ($C_3$) of Y-junctions,
to characterize the orbital structure of triangulene-based nanographenes
and related graphene networks [\onlinecite{TamakiG}].
Such analyses provide useful information about the number and symmetry
properties of the relevant orbitals. However, determining orbital numbers
alone does not fully describe the microscopic evolution of the localized
states as the constituent structures are coupled, nor does it directly
provide the effective hopping amplitudes and Coulomb interactions required
to construct an artificial AGNR. These questions are particularly important
because the resulting electronic properties depend on the sublattice
character of the constituent states, their hybridization,
and the geometry and connectivity of the junction. A real-space description
that tracks the evolution of the localized states is therefore needed to
connect the electronic structure of individual graphene components
to the properties of the assembled artificial quantum material.

Recently synthesized triangulene-based Y-junctions
[\onlinecite{Pascual}] provide an experimentally relevant platform
for investigating these issues. In these structures, three
graphene nanoribbon branches connect through a central
triangulene, forming a junction whose localized electronic states
may receive contributions from both triangulene zero-energy modes
and AGNR end states. Scanning tunneling spectroscopy measurements
have revealed pronounced zero-bias conductance peaks,
demonstrating the presence of low-energy localized
states[\onlinecite{Pascual}]. Understanding their microscopic
origin is important not only for interpreting the experimental
spectra but also for establishing how localized orbitals can be
used to construct more complex graphene architectures. Topological
approaches based on the Zak
phase[\onlinecite{DelplaceP}--\onlinecite{LinKS}] and winding
numbers [\onlinecite{Kariyado}--\onlinecite{LuCH}], within the
framework of bulk-boundary correspondence [\onlinecite{ChenBH}],
can provide information about the existence and number of
interface states. Nevertheless, these approaches do not, by
themselves, fully describe the real-space hybridization between
triangulene zero-energy modes and AGNR end states in a finite
Y-junction. Numerical calculations can determine the electronic
structure of such geometries
[\onlinecite{Maalysheva}--\onlinecite{SanzSO}], but a systematic
analysis of the constituent orbitals and their evolution is needed
to clarify the origin and character of the resulting node states.

In this work, we develop a real-space theoretical framework to
investigate the formation and evolution of localized states in
triangulene-AGNR structures. We consider 3T-GNR/7-AGNR/3T-GNR, and
4T-GNR/7-AGNR/4T-GNR (5T-GNR/7-AGNR/5T-GNR not shown here)
structures, in which triangulene units are connected to AGNR
segments (see Fig. 1(b) and 1(c)). By varying the junction hopping
parameter $t_{es,\alpha}$, we follow the evolution from weakly
coupled constituent states to hybridized junction states and
identify how the triangulene zero-energy modes and AGNR end states
combine to form localized node orbitals. By analyzing the numbers
and sublattice characters of the constituent states, we establish
a counting relation for the node orbitals and examine its
applicability to these three structures. The relation is motivated
by the hybridization physics identified in our previous study of
AGNR heterostructures[\onlinecite{DavidKMT}], but its scope is
assessed here for the triangulene-based junctions investigated in
the present work. We further calculate the tunneling spectra of
the experimentally synthesized structures in Ref
[\onlinecite{Pascual}]  using an extended Anderson model and
Green's-function formalism, thereby connecting our microscopic
analysis with experimentally accessible observables.

Finally, we investigate how the resulting node orbitals can be
used as effective building blocks for constructing artificial
armchair graphene nanoribbons. Here, the term "building blocks"
refers to a real-space effective description in which localized
orbitals and their couplings are identified and incorporated into
a larger electronic structure; it does not imply that the
properties of the assembled system are independent of its geometry
or connectivity. By calculating the effective inter-node hopping
and the relevant intra-node and inter-node Coulomb interactions,
we establish the parameters required to investigate the electronic
properties of these artificial nanoribbons. We show that the
resulting structures support tunable flat subbands near the Fermi
energy, with compact localized states (CLSs) whose degeneracy can
be controlled through the nanoribbon geometry and length
[\onlinecite{Danieli}--\onlinecite{RomerRA}]. We also investigate
their quantum transport properties and demonstrate how the
coupling between localized orbitals influences the resulting
transport behavior. These results provide a microscopic basis for
engineering graphene-based artificial quantum materials through
the controlled arrangement and coupling of localized states.

\begin{figure}[h]
\centering
\includegraphics[trim=1.cm 0cm 1.cm 0cm,clip,angle=0,scale=0.3]{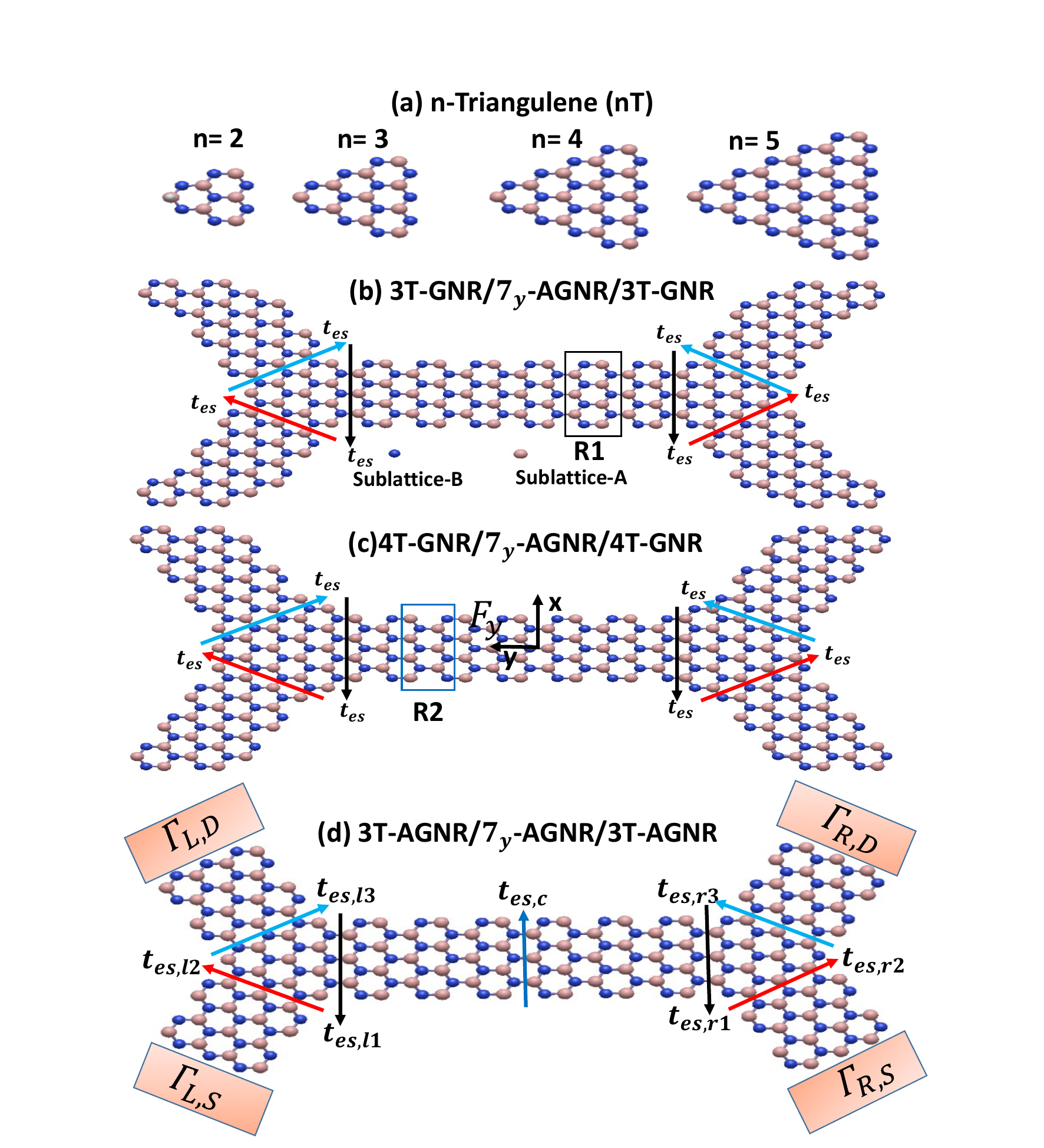}
\caption{Schematic illustrations of the graphene nanostructures
investigated in this work. (a) Atomic structures of
$n$-triangulenes with $n=2$, 3, 4, and 5, illustrating the
sublattice imbalance of $(n-1)$. (b) Schematic of the
3T-GNR/$7_y$-AGNR/3T-GNR nanostructure, consisting of two
3-triangulenes (3T) connected by five graphene nanoribbon (GNR)
segments. The central segment is a 7-atom-wide armchair graphene
nanoribbon (7-AGNR) with the $R_1$ unit-cell (u.c.). The notation
$y$ denotes the bond length between the two 3T structures. The
inter-segment electron hopping parameter $t_{es}$ describes the
coupling between adjacent carbon atoms at the interfaces between
the $n$-triangulene and GNR segments. (c) Schematic of the
4T-GNR/$7_y$-AGNR/4T-GNR nanostructure. In this case, the GNR
segments with the $R_2$ unit-cell structure. An external electric
field $F_y$ is applied along the armchair direction (the $y$
direction) of the central AGNR segment. (d) Device configuration
used for transport calculations. The four outer GNR segments are
replaced by AGNR segments, whose zigzag-terminated ends are
connected to four electrodes. $\Gamma_{L(R),S}$ and
$\Gamma_{L(R),D}$ denote the tunneling rates between the left
(right) source and drain electrodes, respectively, and the
adjacent carbon atoms at the zigzag-terminated edge.}
\end{figure}

\section{Calculation Methodology}
In conventional topological insulators
[\onlinecite{ChangYC}--\onlinecite{HasanMZ}], the bulk consists of
an enormous number of unit cells, making the momentum-space
description a natural framework for characterizing topological
properties. In contrast, atomically precise nanographenes are
finite systems in which boundary effects are no longer negligible.
Consequently, the Zak phase becomes less intuitive, while the
boundary states themselves become the primary objects of interest.
Recent studies have suggested that interface (IF) states can be
understood as descendants of the end states (ESs) of graphene
nanoribbon segments rather than merely as manifestations of bulk
topological invariants [\onlinecite{DavidKMT}]. Motivated by this
real-space picture, we investigate the nanographene structures
shown in Fig.~1(b)--1(d) using a real-space tight-binding
Hamiltonian instead of first-principles density-functional theory
(DFT) calculations with periodic boundary conditions. Although DFT
incorporates electron-electron interactions, its conventional
mean-field formulation remains an effective single-particle theory
[\onlinecite{AnisimovVI}]. Since the present work focuses on the
formation and evolution of localized electronic states governed
primarily by lattice geometry and orbital hybridization, the
one-orbital tight-binding model provides an appropriate and
physically transparent description. The total Hamiltonian is
written as $H=H_0+H_{NG}$ where $H_0$ describes the electrodes and
their coupling to the nanographenes (NGs), while $H_{NG}$ denotes
the Hamiltonian of the nanographene. The electronic structure of
the nanographene is described by a single $p_z$ orbital on each
carbon atom [\onlinecite{NakadaK}--\onlinecite{WakabayashiK2}].

The Hamiltonian $H_0$ is written as:

\begin{small}
\begin{eqnarray}
& & H_0\\ \nonumber & = &\sum_{k,\sigma,\beta} \epsilon_k
a^{\dagger}_{k,\sigma,\beta}a_{k,\sigma,\beta}+
\sum_{k,\sigma,\beta} \epsilon_k b^{\dagger}_{k,\sigma,\beta}b_{k,\sigma,\beta}\\
\nonumber &+&\sum_{\ell,j,k,\sigma,\beta}
V^L_{k,\ell,j,\alpha}d^{\dagger}_{\ell,j,\beta}a_{k,\sigma,\beta}
+\sum_{\ell,j,k,\sigma,\beta}V^R_{k,\ell,j,\beta}d^{\dagger}_{\ell,j,\sigma}b_{k,\sigma,\beta}\\
\nonumber&+& h.c.,
\end{eqnarray}
\end{small}
where the first two terms describe the free electrons in the left
and right $\beta $ electrodes. $a^{\dagger}_{k,\sigma,\beta}$
($b^{\dagger}_{k,\sigma,\beta}$) creates an electron of momentum
$k$ and spin $\sigma$ in the left (right) $\beta$ electrode, with
energy $\epsilon_k$. The terms $V^L_{k,\ell,j,\beta}$
($V^R_{k,\ell,j,\beta}$) describe the coupling between the left
(right) $\beta$ lead and its adjacent atom in the lattice
$(\ell,j)$ site.

The Hamiltonian $H_{NG}$ is given by

\begin{small}
\begin{eqnarray}
H_{NG} &= &\sum_{\ell,j} E_{\ell,j,\sigma} d^{\dagger}_{\ell,j,\sigma}d_{\ell,j,\sigma}\\
\nonumber&-& \sum_{\ell,j}\sum_{\ell',j'} t_{(\ell,j),(\ell', j')}
d^{\dagger}_{\ell,j,\sigma} d_{\ell',j',\sigma} + h.c,
\end{eqnarray}
\end{small}
where $E_{\ell,j}$ is the on-site energy of the $p_z$ orbital at
lattice site $(\ell,j)$. Spin-orbit coupling is neglected because
graphene possesses extremely weak intrinsic spin-orbit interaction
and negligible hyperfine coupling owing to the predominance of
$^{12}$C nuclei with zero nuclear spin
[\onlinecite{AllenMT}--\onlinecite{GerardotBD}].

The operators $d^{\dagger}_{\ell,j,\sigma}$ and
$d_{\ell,j,\sigma}$ create and annihilate an electron with spin
$\sigma$ at site $(\ell,j)$, respectively. The hopping integral
$t_{(\ell,j),(\ell',j')}$ describes electron hopping between
neighboring lattice sites $(\ell,j)$ and $(\ell',j')$. Throughout
this work, the on-site energies are set to $E_{\ell,j}=0$, and the
nearest-neighbor hopping integral is taken as
$t_{(\ell,j),(\ell',j')} = t_{pp,\pi} = t = 2.7$ eV. Sublattices
are labeled A and B in Fig. 1(b) to emphasize the chiral symmetry
of structure[\onlinecite{TamakiG}]. To investigate the evolution
of the localized states across the graphene junctions, a tunable
junction hopping parameter $t_{es,\alpha}$ is introduced to
continuously evolve isolated graphene segments into coupled
Y-junctions in Fig. 1(b) -Fig. 1(d). When $t_{es,\alpha}=0$, the
system is completely decoupled into seven independent components
consisting of two $n$-triangulenes, one central 7-AGNR segment,
and four additional GNR segments. Increasing $t_{es,\alpha}$
continuously modifies the boundary conditions and reveals the
evolution of the end states of the isolated AGNR segments into the
interface states of the coupled system. For the geometry shown in
Fig.~1(d), an additional hopping parameter $t_{es,c}$ is
introduced within the central 7-AGNR segment. In the limit of
$t_{es,c}=0$ and $t_{es,\alpha}=t_{pp,\pi}$, the
3T-AGNR/7-AGNR/3T-AGNR dimer becomes equivalent to two identical
molecules recently synthesized by bottom-up techniques
[\onlinecite{Pascual}]. The effect of a longitudinal electric
field is incorporated through an additional electrostatic
potential $U_y=eF_yy$ added to the on-site energy, where
$F_y=V_y/L_s$, $V_y$ is the applied bias voltage, and $L_s$ is the
length of the nanographene along the $y$ direction.

To investigate the electron transport through the nanographene
junction, the transmission coefficient ${\cal
T}_{SD}(\varepsilon)$ is evaluated using ${\cal
T}_{SD}(\varepsilon) =
4Tr[\Gamma_{L,\beta}(\varepsilon)G^{r}(\varepsilon)\Gamma_{R,\beta}(\varepsilon)G^{a}(\varepsilon)]
$, where $\Gamma_{L(R),\beta}(\varepsilon)$ denotes the coupling
matrix between the nanographene and the left (right) source and
drain electrodes, while $G^{r(a)}(\varepsilon)$ represents the
retarded (advanced) Green's-function matrix of the nanographene
[\onlinecite{SunQF},\onlinecite{Kuo5}]. Within the tight-binding
formalism, both $\Gamma_{L(R),\beta}(\varepsilon)$ and
$G^{r(a)}(\varepsilon)$ are matrix quantities defined in the basis
of atomic $p_z$ orbitals. The expression for
$\Gamma_{L(R),\beta}(\varepsilon)$ is derived from the imaginary
part of the self-energies, denoted as
$\Sigma^r_{L(R),\beta}(\varepsilon)$, and is given by
$\Gamma_{L(R),\beta}(\varepsilon)=-\text{Im}(\Sigma^r_{L(R),\beta}(\varepsilon))=\pi\sum_k
|V^{L(R)}_{k,\ell,j, \beta}|^2\delta(\varepsilon-\epsilon_k)$.The
indices $(j,\ell,\beta)$ identify the contacted atomic or node
sites and the corresponding electrode. The coupling matrix
elements $V^{L(R)}_{k,j,\ell,\beta}$ determine the
electrode-nanostructure coupling and enter the corresponding
self-energies $\Sigma^r_{L(R),\beta}(\varepsilon)$. For the
four-terminal configuration, the subscript of ${\cal
T}_{L(R),\beta=S,D}(\varepsilon)$ specifies the source- and
drain-electrode combinations considered in calculating the
transmission. These combinations are included in the transmission
formulation for the geometry illustrated in Fig. 1(d). For
simplicity, we adopt the wide-band limitation for the tunneling
rates, $\Gamma_{L(R),\beta}(\varepsilon)$, assuming that these
rates are energy-independent, and denote them as
$\Gamma_{L(R),\beta}$. The energy spectra, charge-density
distributions, and transmission coefficients are obtained by
numerically diagonalizing the tight-binding Hamiltonian and
evaluating the corresponding Green's functions. All numerical
calculations were performed using self-developed Fortran
codes[\onlinecite{Kuo5}].

\section{Results and Discussion}

\subsection{Evolution of in-gap states in $n$T-GNR/$7_y$-AGNR/$n$T-GNR
structures} To elucidate the interface properties between
$n$-triangulenes  and graphene nanoribbons (GNRs), we first
calculate the energy levels of the 3T-GNR/$7_y$-AGNR/3T-GNR
structure as functions of the junction hopping parameter $t_{es}$,
as illustrated in Fig.~1(b), for different lengths of the central
$7_y$-AGNR segment. Here, the subscript $y$ denotes the length of
the 7-AGNR segment in units of $R_1$, where $y=4$, $5$, $6$, and
$7$ correspond to Figs.~2(a)-2(d), respectively. When $t_{es}=0$,
the system is completely decoupled into seven isolated graphene
segments, giving rise to ten in-gap states: four zero-energy modes
contributed by the two triangulenes (3T), two zero-energy modes
originating from the central 7-AGNR segment, and four zero-energy
modes associated with the four outer GNR segments. Unlike the
central 7-AGNR segment, which possesses two zigzag termini, each
outer GNR segment contains only a single zigzag edge. As $t_{es}$
increases continuously from 0 to $t$, four energy levels denoted
by $\Sigma_{AB,C}$ and another four denoted by $\Sigma_{AB,V}$
move away from the charge-neutral point (CNP), resulting in two
in-gap states located inside the bulk energy gap.

Figure~2(a) shows the evolution of the four $\Sigma_{AB,C}$ and
four $\Sigma_{AB,V}$ energy levels, which arise from the
hybridization of these eight in-gap states. The subscripts $C$ and
$V$ denote the conduction- and valence-side energy levels,
respectively. As the length of the 7-AGNR segment increases, the
four $\Sigma_{AB,C}$ ($\Sigma_{AB,V}$) branches gradually merge
into two nearly degenerate branches, as illustrated in Fig.~2(d).
These remaining in-gap energy levels become nearly independent of
$t_{es}$ for $t_{es}\ge 0.2\,t$. They are labeled $\Sigma_{IF,C}$
and $\Sigma_{IF,V}$, corresponding to the bonding and antibonding
states formed by the left and right interface (IF) states,
$\psi_{IF,A,0}$ and $\psi_{IF,B,0}$, respectively. The left
(right) interface state can be approximately expressed as $
\psi_{IF,A(B),0}=
(\phi_{A(B),1}+\phi_{A(B),2}+\phi_{A(B),3})/\sqrt{3}$, indicating
that it is primarily composed of an equal-weight linear
combination of the three end states (ESs) with sublattice-$A$
($B$) chirality. This wave function contains one orbital node at
each left (right) Y-junction, consistent with the
single-orbital-node solution dictated by the $C_3$ symmetry of
three semi-infinite graphene-arm junctions [\onlinecite{TamakiG}].
The end states $\phi_{A(B)}$ of the isolated AGNR segments are
discussed in Appendix A (see Fig.~A.1). At $t_{es}=t$, the energy
splitting between $\Sigma_{IF,C}$ and $\Sigma_{IF,V}$ determines
the effective electron hopping strength $t_{SL}$ between the left
and right interface states.

The dependence of $t_{SL}$ on the length of the central 7-AGNR
segment is presented in Appendix B (see Fig.~B.1), together with
the calculated intra-interface-state and inter-interface-state
Coulomb interactions. The effective hopping strength decreases
exponentially with increasing $y$, yielding $t_{SL}\approx 29$~meV
and $17$~meV for $y=4$ and $5$, respectively. This wide tunability
of $t_{SL}$ provides an important degree of freedom for
engineering the electronic subbands of artificial graphene-based
quantum materials.

\begin{figure}[h]
\centering
\includegraphics[trim=1.cm 0cm 1.cm 0cm,clip,angle=0,scale=0.3]{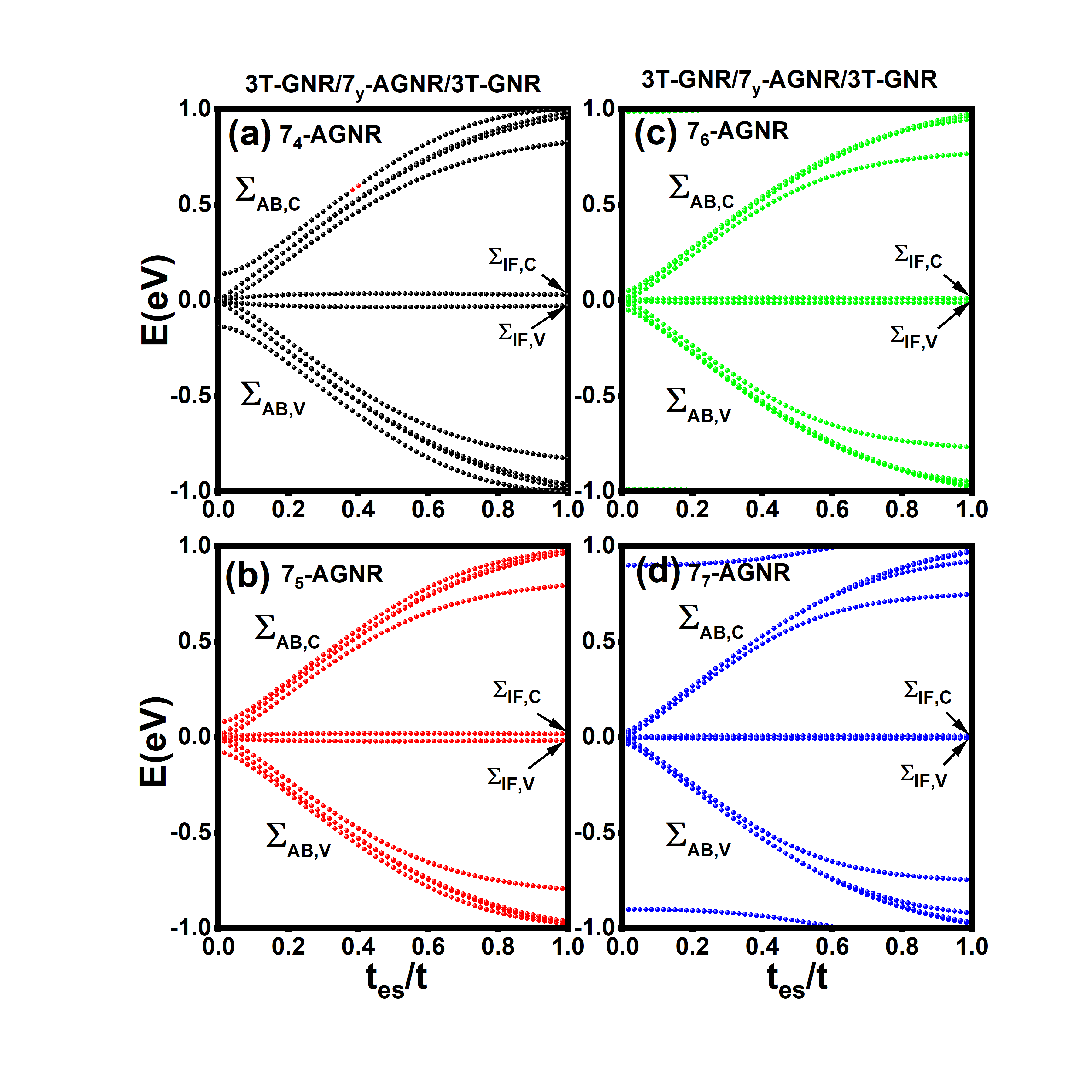}
\caption{Energy levels of the 3T-GNR/$7_y$-AGNR/3T-GNR
heterostructure as functions of $t_{es}$ for different 7-AGNR
segment lengths: (a) $7_4$-AGNR, (b) $7_5$-AGNR, (c) $7_6$-AGNR,
and (d) $7_7$-AGNR. Here, the subscript $y$ denotes the length of
the 7-AGNR segment with $R_1 = 3a_{cc}$ per unit cell (u.c.),
where $a_{cc}$ is the carbon--carbon bond length.}
\end{figure}

In the 3T-GNR/$7_y$-AGNR/3T-GNR structure, each 3-triangulene is
connected to three armchair GNR branches through a Y junction. A
natural question is how this picture changes when the
3-triangulene is replaced by a 4-triangulene, which possesses
three zero-energy modes instead of two. In particular, we ask how
many localized orbitals survive at each Y junction after coupling
to the three armchair GNR segments.

Figure~3 presents the energy levels of the
4T-GNR/$7_y$-AGNR/4T-GNR dimer as functions of the junction
hopping parameter $t_{es}$ for different central 7-AGNR lengths,
$y=8$, 7, 6, and 5. At $t_{es}=0$, the system consists of seven
isolated graphene segments and exhibits 18 in-gap states,
including six zero-energy modes contributed by the two
4-triangulenes, four zero-energy modes originating from the
central 7-AGNR segment with the $R_2$-type unit cell, and eight
zero-energy modes associated with the four outer GNR segments,
which also possess $R_2$-type zigzag edge structures. As $t_{es}$
increases from 0 to $t$, twelve of these zero-energy modes move
away from the CNP because zero-energy modes with opposite
chiralities hybridize at the junction sites.

For the longest central segment ($y=8$), the hybridized energy
levels are labeled by $\Sigma_{AB,C(V),i=1,2,3}$ in Fig.~3(a). The
branches $\Sigma_{AB,C(V),1}$ and $\Sigma_{AB,C(V),3}$ remain
doubly degenerate over a wide range of $t_{es}$, whereas the
degeneracy of $\Sigma_{AB,C(V),2}$ is lifted when
$t_{es}\gtrsim0.5\,t$. As the central 7-AGNR segment becomes
shorter, the evolution of the remaining in-gap states changes
significantly. For $y=7$, two zero-energy modes split from the
fourfold-degenerate manifold, while the corresponding energy
splitting remains relatively small at $t_{es}=t$, as shown in
Fig.~3(b). For $y=6$ and 5, the splitting becomes sufficiently
large to be clearly resolved, giving rise to the bonding and
antibonding interface states labeled $\Sigma_{IF,C}$ and
$\Sigma_{IF,V}$ in Figs.~3(c) and 3(d), respectively.

The evolution of $\Sigma_{IF,C}$ and $\Sigma_{IF,V}$ provides
important insight into their physical origin. Their energy
separation is largest at $t_{es}=0$ and decreases continuously as
$t_{es}$ approaches $t$, reaching a minimum in the fully coupled
structure. This behavior indicates that these two states originate
from the end states of the central 7-AGNR segment. At $t_{es}=0$,
the wave functions $\psi_{A,1}$ and $\psi_{B,1}$ are entirely
confined within the central 7-AGNR segment, leading to the
strongest wave-function overlap and the largest energy splitting.
As the junction coupling increases, the end-state wave functions
gradually leak into the neighboring triangulene sites with the
same chirality, thereby reducing their mutual overlap and
decreasing the bonding-antibonding splitting. The remaining six
in-gap states are localized interface states consisting of three
$A$-chirality orbitals at the left Y junction and three
$B$-chirality orbitals at the right Y junction, which are denoted
collectively by $\Sigma_{IF,0}$ and $\Sigma_{IF,C(V)}$.

\begin{figure}[h]
\centering
\includegraphics[trim=1.cm 0cm 1.cm 0cm,clip,angle=0,scale=0.3]{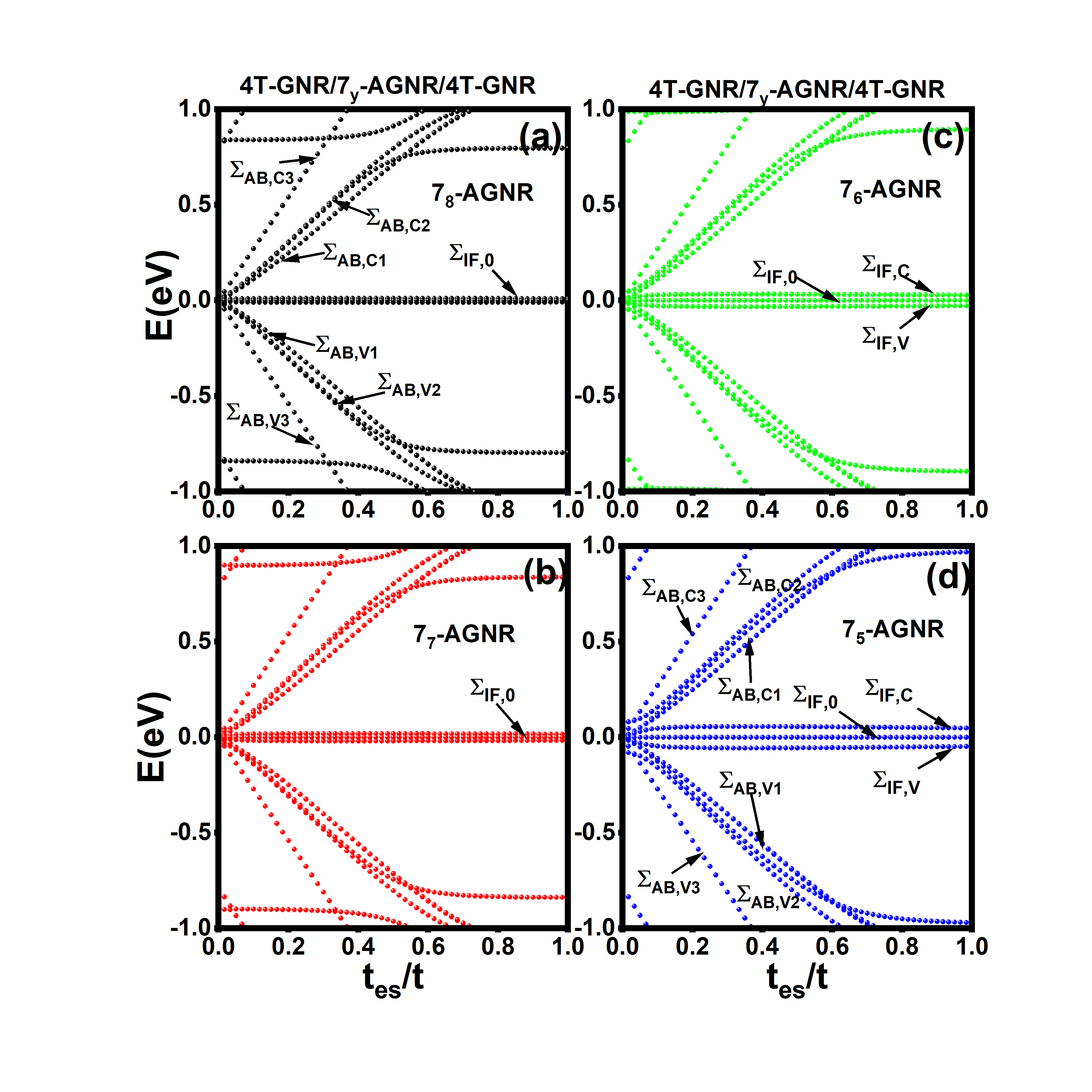}
\caption{ Energy levels of the 4T-GNR/$7_y$-AGNR/4T-GNR
heterostructure as functions of $t_{es}$ for different 7-AGNR
segment lengths: (a) $7_8$-AGNR, (b) $7_7$-AGNR, (c) $7_6$-AGNR,
and (d) $7_5$-AGNR.}
\end{figure}

The 4T-GNR/$7_y$-AGNR/4T-GNR nanostructure therefore supports six
localized interface states within the bulk energy gap, as shown in
Fig.~3. To further identify their number and chirality, we apply a
longitudinal electric field along the armchair direction of the
central 7-AGNR segment. The resulting Stark effect provides an
effective spectroscopic probe for distinguishing the surviving
interface states.

Figures~4(a)-4(d) present the calculated energy levels of the
4T-GNR/$7_y$-AGNR/4T-GNR structure as functions of the applied
voltage $V_y$ for $y=8$, 7, 6, and 5 at $t_{es}=t$. Two distinct
types of Stark shifts are observed. The delocalized bulk states
($E_C$ and $E_V$) move toward the CNP, corresponding to a Stark
red shift, whereas the localized interface states shift away from
the CNP, corresponding to a Stark blue shift. The interface-state
energy levels, labeled $\Sigma_{A,i=1,2,3}$ and
$\Sigma_{B,i=1,2,3}$, exhibit an approximately linear dependence
on the applied voltage. Owing to their different spatial
distributions, the three sublattice-$A$ chirality and three
sublattice-$B$ chirality interface states become well resolved
under the electric field. These results demonstrate that the
longitudinal Stark effect serves as an effective spectroscopic
tool for resolving the degeneracy and chirality of the zero-energy
interface states.

\begin{figure}[h]
\centering
\includegraphics[trim=1.cm 0cm 1.cm 0cm,clip,angle=0,scale=0.3]{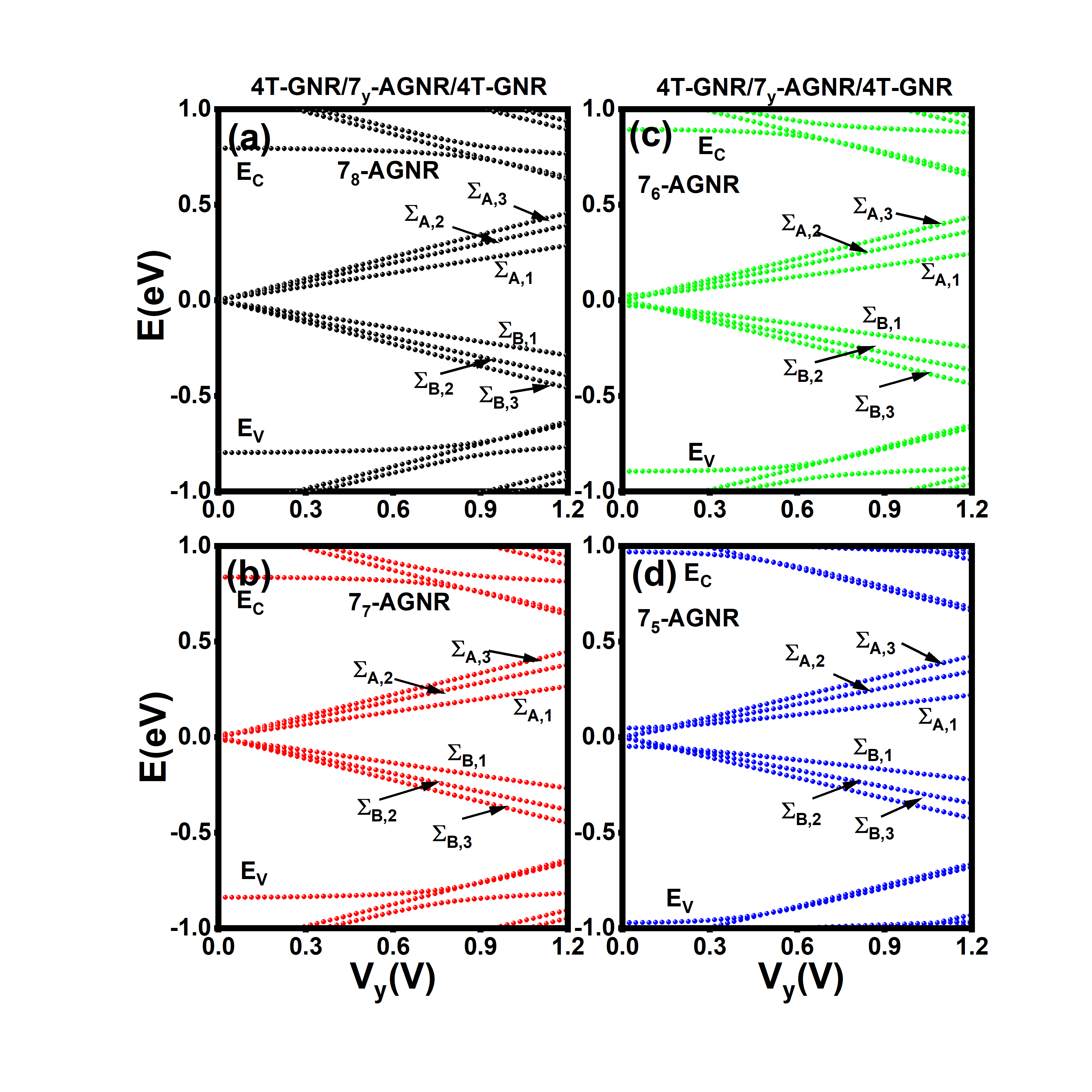}
\caption{Energy levels of the 4T-GNR/$7_y$-AGNR/4T-GNR
heterostructure as functions of the applied voltage $V_y$ at
$t_{es}=t$ for different 7-AGNR segment lengths: (a) $7_8$-AGNR,
(b) $7_7$-AGNR, (c) $7_6$-AGNR, and (d) $7_5$-AGNR.}
\end{figure}

\subsection{Charge densities of $n$T-GNR/$7$-AGNR/$n$T-GNR structures}

To clarify the physical origin of the in-gap states identified in
Figs.~2 and 3, we examine the evolution of the charge-density
distributions of the interface state $\Sigma_{IF,C}$ as the
junction hopping parameter $t_{es}$ is varied. Figure~5(a)-5(e)
shows the charge densities of $\Sigma_{IF,C}$ for the
3T-GNR/$7_6$-AGNR/3T-GNR structure at $t_{es}=0.1\,t$, $0.2\,t$,
$0.5\,t$, $0.7\,t$, and $t$, respectively. The corresponding
results for the 4T-GNR/$7_6$-AGNR/4T-GNR structure are presented
in Figs.~5(f)-5(i) for $t_{es}=0.1\,t$, $0.2\,t$, $0.5\,t$, and
$t$. The continuous evolution of these charge-density
distributions directly reveals how the interface states emerge
from the ESs of the isolated GNR segments.

For the 3T-GNR/$7_6$-AGNR/3T-GNR structure, the charge densities
shown in Figs.~5(a) and 5(b) remain almost completely localized on
the zigzag edges of the three GNR segments when $t_{es}=0.1\,t$
and $0.2\,t$. The probability density on the triangulene is
negligibly small, indicating that the coupling between the
triangulene and the GNR segments is still weak. In this regime,
the outer GNR end states, $\phi_{A(B),1}$ and $\phi_{A(B),2}$,
carry larger probability weights than the end state of the central
7-AGNR segment, $\phi_{A(B),3}$. As the junction coupling
increases to $t_{es}=0.5\,t$, the three end states acquire nearly
identical probability weights, indicating that they contribute
equally to the interface state. A further increase to
$t_{es}=0.7\,t$ allows the wave function to penetrate into the
triangulene through the corresponding sublattice-$A$ ($B$) sites.
Finally, at $t_{es}=t$, the interface state becomes an
equal-weight superposition of the three GNR end
states,$\psi_{A(B),0}=\frac{\phi_{A(B),1}+\phi_{A(B),2}+\phi_{A(B),3}}{\sqrt{3}}$,
demonstrating that the localized interface orbital originates from
the coherent hybridization of the three ESs attached to the Y
junction.

A similar evolution is observed for the 4T-GNR/$7_6$-AGNR/4T-GNR
structure, although its physical origin is different. As shown in
Figs.~5(f) and 5(g), the charge densities remain localized at the
zigzag edges of both the outer GNR segments and the central 7-AGNR
segment for $t_{es}=0.1\,t$ and $0.2\,t$, with the dominant
contribution coming from the zigzag termini of the central 7-AGNR
segment. As $t_{es}$ increases to $0.5\,t$, the probability
densities on the outer GNR segments are substantially suppressed,
while the wave functions become increasingly concentrated around
the central 7-AGNR and neighboring triangulene sites. In the fully
coupled limit ($t_{es}=t$), the interface state is primarily
described by the hybridization of the two end states, $\psi_{A,1}$
and $\psi_{B,1}$, located at the opposite zigzag termini of the
central 7-AGNR segment. Their wave functions extend into the
adjacent triangulene through carbon atoms belonging to the same
sublattice, illustrating the strong coupling between the central
ribbon and the triangulene building blocks.

The charge-density evolution shown in Fig.~5 provides a direct
real-space interpretation of the interface states. Combined with
the energy spectra presented in Figs.~2 and 3, these results
demonstrate that the 3T-GNR/7-AGNR/3T-GNR dimer supports one
localized orbital at each Y junction, whereas the
4T-GNR/7-AGNR/4T-GNR dimer supports three localized orbitals at
each junction. Moreover, the junction orbitals on the left and
right sides always possess opposite chiralities (sublattice-A and
sublattice-B), preserving the intrinsic sublattice (chiral)
symmetry of the graphene nanostructures shown in Figs.~1(b) and
1(c). These localized junction orbitals constitute the elementary
quantum building blocks from which the artificial quantum
materials discussed in the following sections are constructed.

By analyzing the numbers and sublattice characters of the
constituent states for 3T-GNR/7-AGNR/3T-GNR, 4T-GNR/7-AGNR/4T-GNR,
and 5T-GNR/7-AGNR/5T-GNR (not shown here), we establish a counting
relation for the node orbitals and examine its applicability to
these three structures. The relation is motivated by the
hybridization physics identified in our previous study of AGNR
heterostructures [\onlinecite{DavidKMT}], but its scope is
assessed here for the triangulene-based honeycomb structures with
sublattice-A(B) chirality [\onlinecite{TamakiG}]. The number and
chirality of the node orbitals, $N_{node,\delta}$, can be
determined by $N_{node,\delta}=|N_{es,t,A(B)}-N_{tri,0,B(A)}|$,
where $N_{es,t}$ represents the total number of AGNR end states
contributed by the three AGNR arms at the junction, and
$N_{tri,0}$ denotes the number of zero-energy modes of the
attached triangulene. This relation, referred to here as the
junction-orbital counting rule, provides a simple real-space
criterion for predicting the number of localized interface
orbitals without solving the complete electronic structure of the
coupled system. Furthermore, the chirality of the junction
orbitals (determined by charge density in sublattice-A(B)) is
determined by the dominant constituent: when $N_{es,t}>N_{tri,0}$,
the junction orbitals inherit the chirality of the GNR end states
($\delta=A(B)$), whereas for $N_{tri,0}>N_{es,t}$ they inherit the
chirality of the triangulene zero-energy modes ($\delta=B(A)$).
Consequently, this counting rule predicts a single node orbital
for the left (right) 5T-GNR/7-AGNR (7-AGNR/5T-GNR) Y-junction,
originating from the zero-energy mode of the 5T triangulene with
sublattice-B (sublattice-A) chirality.

\begin{figure}[h]
\centering
\includegraphics[trim=1.cm 0cm 1.cm 0cm,clip,angle=0,scale=0.25]{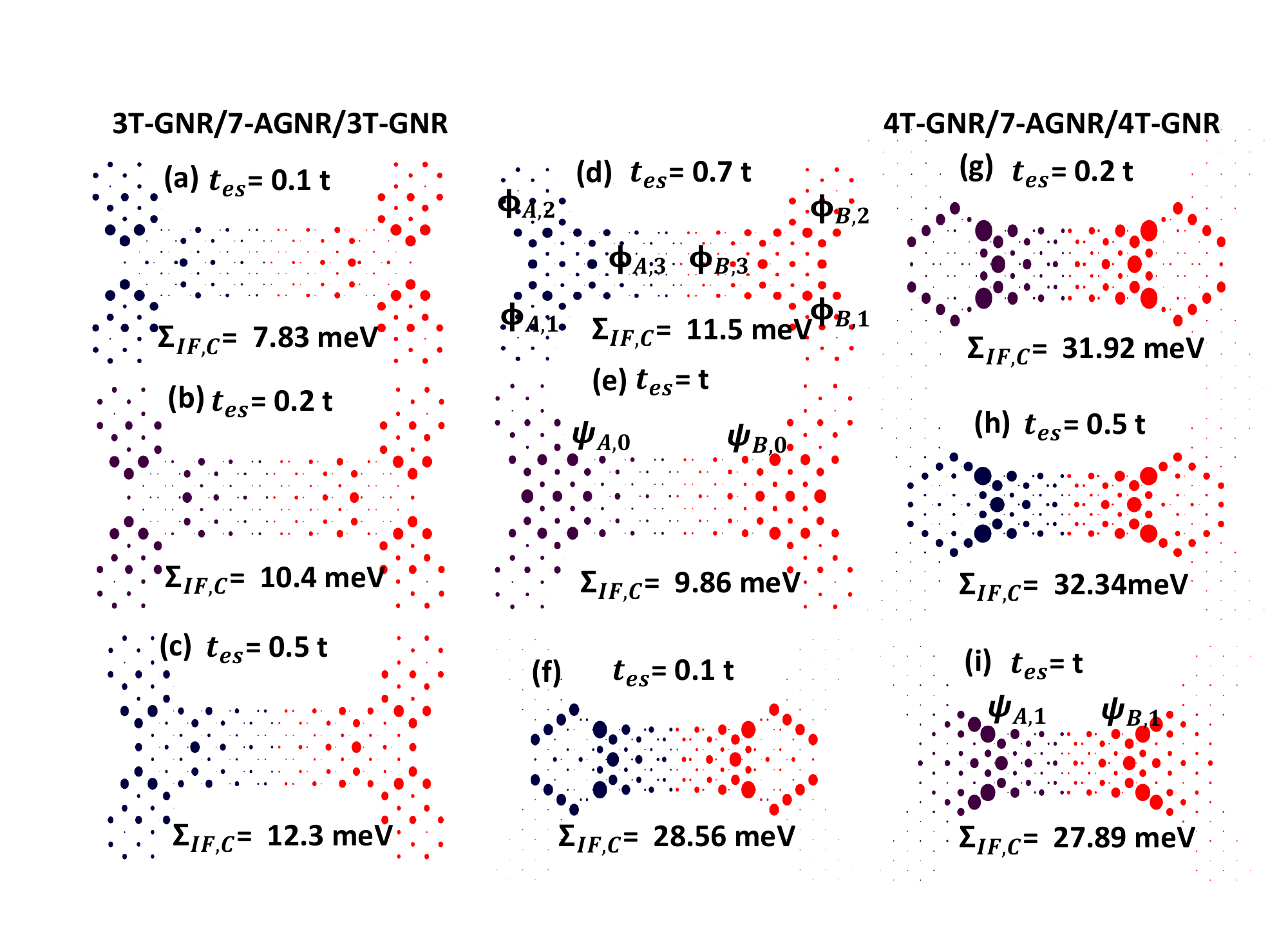}
\caption{(a)-(e) Charge density distributions of the
3T-GNR/$7_6$-AGNR/3T-GNR heterostructure for different values of
$t_{es}$: (a) $t_{es}=0.1~t$, (b) $t_{es}=0.2~t$, (c)
$t_{es}=0.5~t$, (d) $t_{es}=0.7~t$, and (e) $t_{es}=t$. (f)-(i)
Charge density distributions of the 4T-GNR/$7_6$-AGNR/4T-GNR
heterostructure for different values of $t_{es}$: (f)
$t_{es}=0.1~t$, (g) $t_{es}=0.2~t$, (h) $t_{es}=0.5~t$, and (i)
$t_{es}=t$.}
\end{figure}

\subsection{3T-AGNR/$7_y$-AGNR/3T-AGNR dimers}

The results presented in the previous subsections establish the
number and chirality of the localized orbitals formed at the
triangulene Y junctions. Although the nanographene dimers shown in
Figs.~1(b) and 1(c) provide a clear understanding of the
microscopic origin of these interface states, their terminal GNR
segments cannot be directly extended into larger artificial
lattices. To construct scalable graphene-based artificial quantum
materials, we therefore replace the four outer GNR segments with
AGNR segments, leading to the
3T-$7_w$-AGNR/$7_y$-AGNR/3T-$7_w$-AGNR dimer shown in Fig.~1(d).
This geometry can be naturally connected to neighboring
triangulene units and subsequently extended into a two-dimensional
honeycomb network. The resulting structures correspond to the
Anthracene-Aza-3-triangulene (AAT), Bisanthene-Aza-3-triangulene
(BAT), and Teranthene-Aza-3-triangulene (TAT) dimers for $y=2$
($w=1$), $y=4$ ($w=2$), and $y=6$ ($w=3$), respectively.

Figure~6(a) presents the calculated energy levels of the
3T-$7_2$-AGNR/$7_4$-AGNR/3T-$7_2$-AGNR dimer (BATD) as functions
of the junction hopping parameter $t_{es}$ in the absence of
electrodes. Six in-gap states appear around the CNP, labeled by
$\Sigma_{0}$, $\Sigma_{C1}$, $\Sigma_{C2}$, $\Sigma_{V1}$, and
$\Sigma_{V2}$. The two $\Sigma_{0}$ states originate from weakly
coupled zero-energy modes and therefore remain nearly degenerate
throughout the variation of $t_{es}$. Compared with the spectra
shown in Fig.~2, the BATD structure exhibits four additional
in-gap states because the four outer AGNR segments with $R_1$ unit
cells each contribute an additional end state.

The transport properties associated with these in-gap states are
illustrated by the transmission coefficient ${\cal
T}_{SD}(\varepsilon)$ shown in Figs.~6(b)-6(e) for different
lengths of the central $7_y$-AGNR segment: $y=4$, 5, 6, and 7. The
transmission peak associated with $\Sigma_{0}$ reaches
approximately two and remains nearly independent of the central
AGNR length, demonstrating the existence of two robust zero-energy
modes in the BATD structure. This behavior is markedly different
from that of the interface states $\Sigma_{IF,C}$ and
$\Sigma_{IF,V}$ shown in Fig.~2, whose energy splitting depends
sensitively on the coupling between the two Y junctions. By
contrast, the energy separation between $\Sigma_{C(V)1}$ and
$\Sigma_{C(V)2}$ decreases continuously as the length of the
central 7-AGNR segment increases, indicating that these states are
strongly influenced by the inter-junction coupling.

The origin of these transmission spectra can be understood from
the corresponding charge-density distributions of $\Sigma_{0}$,
$\Sigma_{C1}$, and $\Sigma_{C2}$, shown in Figs.~6(f)-6(h) for the
3T-$7_2$-AGNR/$7_4$-AGNR/3T-$7_2$-AGNR dimer. The nearly
degenerate $\Sigma_{0}$ states ($E\approx1.1$~meV) are primarily
composed of the end states of the four outer AGNR segments,
$\varphi_{B1}$, $\varphi_{B2}$, $\varphi_{A1}$, and
$\varphi_{A2}$. In contrast, the wave functions of $\Sigma_{C1}$
and $\Sigma_{C2}$ contain contributions from both the outer AGNR
end states and the interface orbitals $\psi_{A0}$ and $\psi_{B0}$
[see Fig.~5(e)]. These charge-density distributions indicate that
the in-gap states of the BATD are constructed from three localized
orbitals at each Y junction, namely
$(\varphi_{B1},\varphi_{B2},\psi_{A0})$ on the left and
$(\varphi_{A1},\varphi_{A2},\psi_{B0})$ on the right.

The two zero-energy states can be written approximately as $
\Psi_{B(A),0}
=\frac{\varphi_{B1(A1)}-\varphi_{B2(A2)}}{\sqrt{2}}$, which are
highly localized on the outer AGNR segments. Because the overlap
between $\Psi_{A,0}$ and $\Psi_{B,0}$ is extremely small, the
resulting bonding-antibonding splitting is negligible.
Consequently, the $\Sigma_{0}$ energy levels and their
corresponding transmission peaks remain almost unchanged as the
length of the central $7_y$-AGNR segment varies. By contrast,
$\Sigma_{C1(V1)}$ and $\Sigma_{C2(V2)}$ involve the interface
orbitals $\psi_{A0}$ and $\psi_{B0}$ and therefore exhibit a
strong dependence on the central AGNR length. The transmission
peak associated with $\Sigma_{0}$ further demonstrates that
long-distance coherent tunneling can occur directly between
$\varphi_{B(A),1}$ and $\varphi_{B(A),2}$ without the assistance
of the interface orbitals $\psi_{A(B),0}$.

\begin{figure}[h]
\centering
\includegraphics[trim=1.cm 0cm 1.cm 0cm,clip,angle=0,scale=0.25]{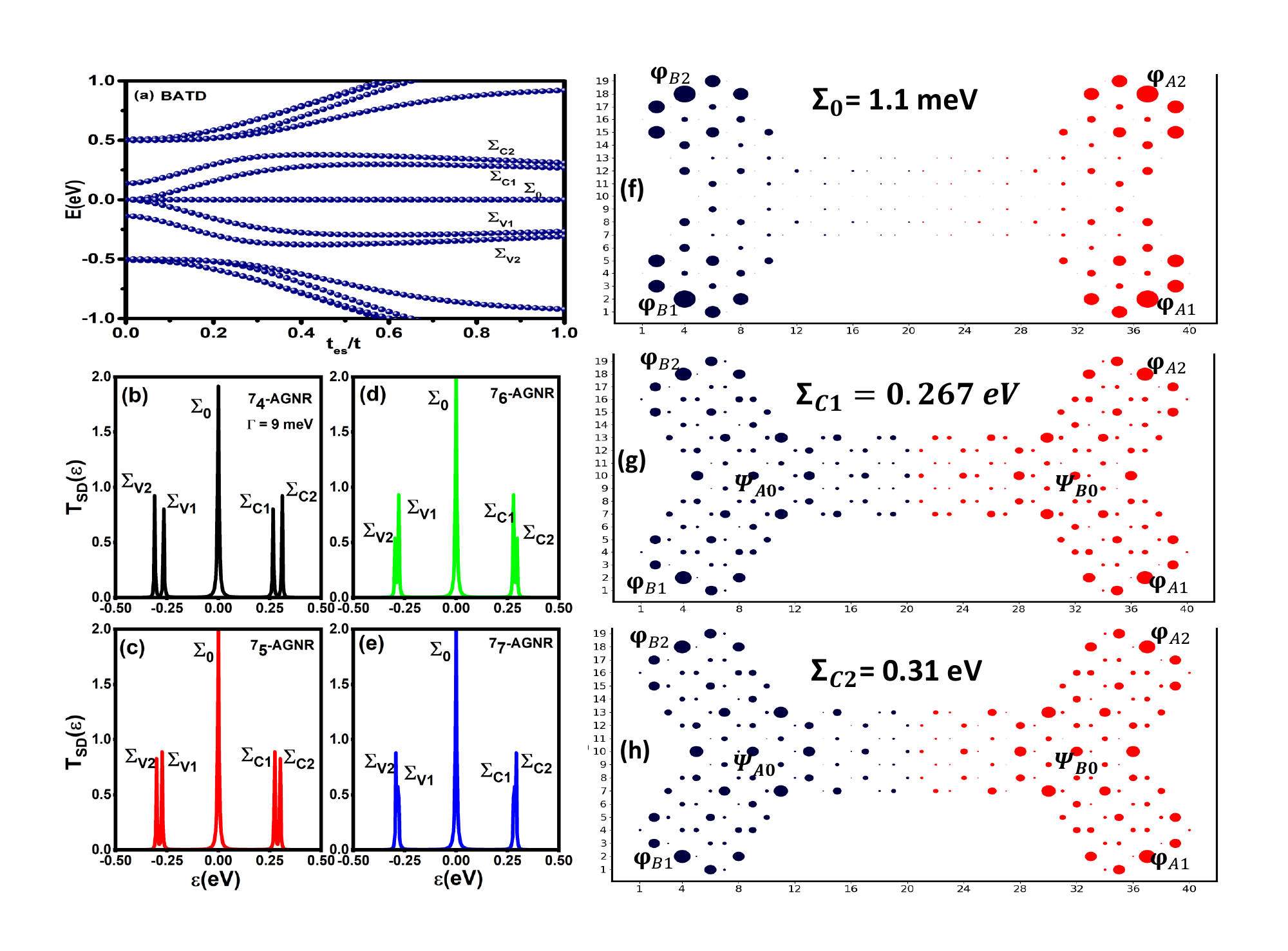}
\caption{(a) Energy levels of the BAT dimer (BATD) as functions of
$t_{es}$ for the $7_4$-AGNR segment. (b)-(e) Transmission
coefficient ${\cal T}_{SD}(\varepsilon)$ of the BATD for different
$7_y$-AGNR segment lengths at
$\Gamma_{L,S(D)}=\Gamma_{R,S(D)}=\Gamma=9$~meV. (f)-(h) Charge
density distributions corresponding to $\Sigma_{C0}=1.1$~meV,
$\Sigma_{C1}=0.267$~eV, and $\Sigma_{C2}=0.31$~eV for the
3T-$7_2$-AGNR/$7_4$-AGNR/3T-$7_2$-AGNR heterostructure (BATD).}
\end{figure}

\subsection{Orbitals and charge densities of AAT and BAT molecules}

The robust zero-bias differential conductance peaks observed in
STM measurements of AAT and BAT molecules [\onlinecite{Pascual}]
indicate the presence of stable zero-energy states. Understanding
the molecular orbital origin of these states is therefore
essential. To investigate the orbital structures of AAT and BAT
molecules theoretically, we decouple the central 7-AGNR segment by
setting $t_{es,C}=0$ while maintaining $t_{es,\alpha,i} = t$ for
the connections between the triangulene and the AGNR segments in
AATD and BATD [Fig.~1(d)]. Under this condition, the left and
right AAT (BAT) molecules become independent and possess opposite
chiralities. Figures~7(a)-7(d) and 7(e)-7(h) present the
charge-density distributions of the four energy levels closest to
the CNP for the left AAT and BAT molecules, respectively, under a
small applied voltage $V_y$. The Stark shift induced by $V_y$
lifts the degeneracy of the zero-energy modes and enables their
individual identification.

Compared with the charge distributions of the dimers shown in
Fig.~6, the isolated AAT and BAT molecules contain an additional
end state $\varphi_{B3}$ originating from the central 7-AGNR
segment. The low-energy electronic structure can therefore be
described by four localized orbitals, $\psi_{A,0}$,
$\varphi_{B1}$, $\varphi_{B2}$, and $\varphi_{B3}$, with on-site
energies $\epsilon=0$. The only nonzero coupling is the hopping
parameter $t_{AB}$ between the node orbital $\psi_{A,0}$ and the
three GNR end states $\varphi_{B,i}$ ($i=1,2,3$). The
corresponding effective Hamiltonian yields four eigenvalues at
$V_y=0$:

\[
\Sigma_{\pm}=\pm\sqrt{3}t_{AB}, \qquad \Sigma_{0,1}=0,\qquad
\Sigma_{0,2}=0 .
\]

The two zero-energy eigenfunctions are

\[
\Psi^0_{1}
=\frac{\varphi_{B1}+\varphi_{B2}-2\varphi_{B3}}{\sqrt{6}},
\]

and

\[
\Psi^0_{2} =\frac{\varphi_{B1}-\varphi_{B2}}{\sqrt{2}},
\]

respectively. These states are completely decoupled from
$\psi_{A,0}$ because of destructive interference among the three
GNR end states. The two finite-energy eigenstates are given by

\[
\Psi_{+} =\frac{\sqrt{3}\psi_{A0}
+\varphi_{B1}+\varphi_{B2}+\varphi_{B3}} {\sqrt{6}},
\]

and

\[
\Psi_{-} =\frac{\sqrt{3}\psi_{A0}
-\varphi_{B1}-\varphi_{B2}-\varphi_{B3}} {\sqrt{6}},
\]
which correspond to the bonding and antibonding combinations
between the node orbital and the three GNR end states.

From the calculated energy splitting, we obtain $t_{AB}=0.48$~eV
and $0.215$~eV for AAT and BAT molecules, respectively. These
values indicate strong hybridization between the node orbital and
the surrounding GNR end states. Furthermore, the two zero-energy
states exhibit negligible contribution from $\psi_{A,0}$ and
therefore remain highly localized on the zigzag edge states of the
AGNR segments. Their strong localization and associated Coulomb
interactions for double occupation provide a natural explanation
for the robust zero-bias differential conductance peaks observed
in STM measurements of AAT and BAT molecules on copper substrates.
Since these zero-energy states do not contain the node orbital
$\psi_{A,0}$, the influence of the nitrogen substitution at the
central site of 3-triangulene on the STM tunneling current is
expected to be negligible.

\begin{figure}[h]
\centering
\includegraphics[angle=0,scale=0.25]{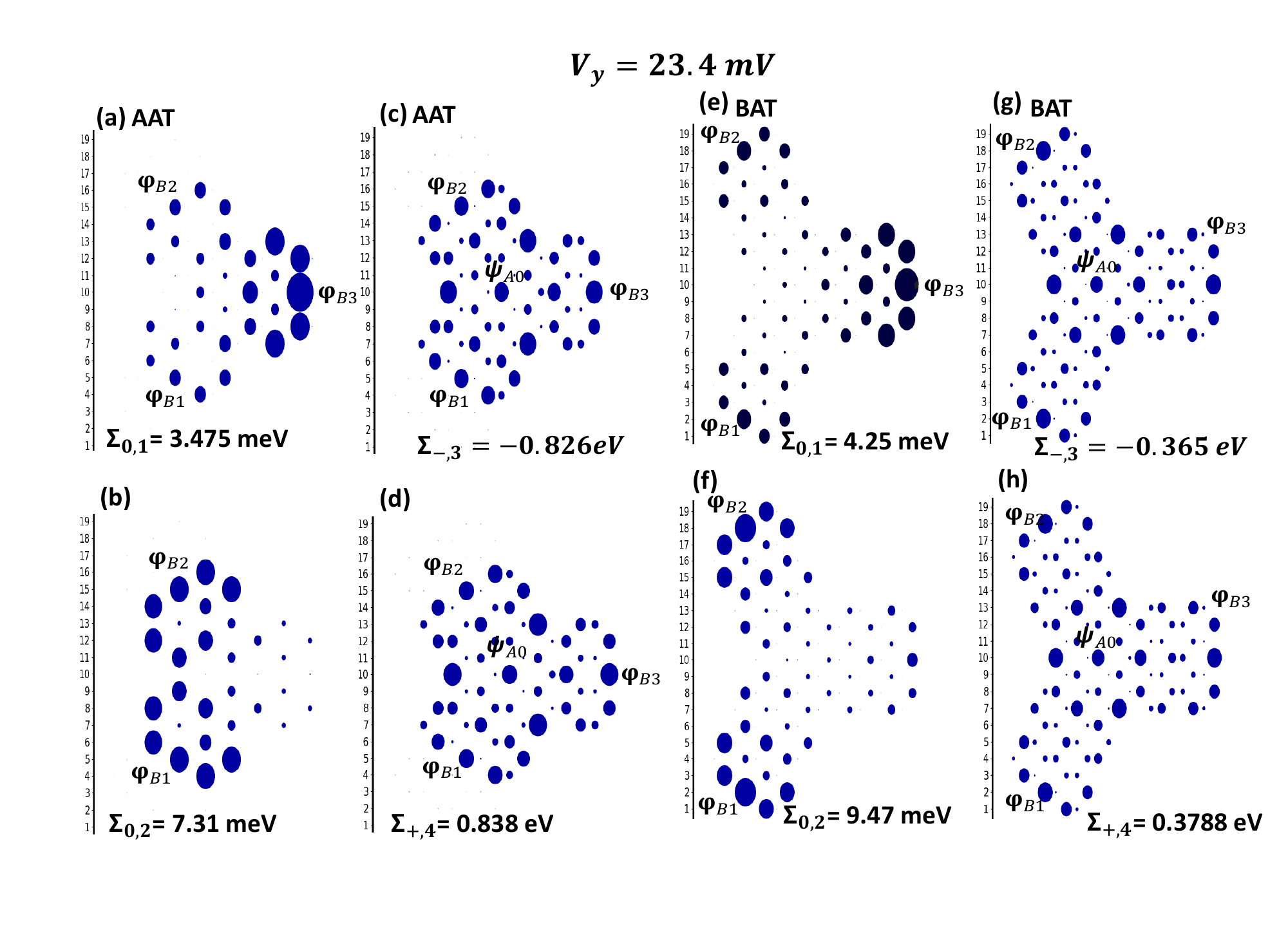}
\caption{(a)-(d) Charge density distributions corresponding to
$\Sigma_{0,1}=3.475$~meV, $\Sigma_{0,2}=7.31$~meV,
$\Sigma_{-,3}=-0.826$~eV, and $\Sigma_{+,4}=0.838$~eV for the left
AAT nanostructure. (e)-(h) Charge density distributions
corresponding to $\Sigma_{0,1}=4.25$~meV, $\Sigma_{0,2}=9.47$~meV,
$\Sigma_{-,3}=-0.365$~eV, and $\Sigma_{+,4}=0.3788$~eV for the
left BAT nanostructure.}
\end{figure}

\subsection{Tunneling current of STM through AAT and BAT molecules on metallic substrates}

The transmission coefficients ${\cal T}_{SD}(\varepsilon)$
presented in Fig.~6 are calculated within a single-particle
framework. However, a quantitative description of tunneling
transport in the Coulomb blockade regime requires explicit
consideration of electron-electron interactions, which remains a
challenging problem for nanoscale systems
[\onlinecite{SolsF}--\onlinecite{ZhangJain}]. When on-site Coulomb
interactions are included in the Hamiltonian of Eq.~(2), existing
theoretical approaches are generally restricted to mean-field
approximations and therefore cannot accurately describe correlated
transport phenomena such as Coulomb blockade or Kondo physics
[\onlinecite{JacobsePH}]. Since the present work focuses on the
low-energy zero modes near the CNP, we employ an extended Anderson
model that incorporates both intra-level and inter-level Coulomb
interactions to investigate the STM tunneling properties of the
two zero-energy modes $\Sigma_0$ in AAT and BAT molecules.

The effective Hamiltonian is written as
$H_{eff}=H_{SD}+H_{\Sigma_0}$, where $H_{SD}$ describes the source
and drain electrodes  and their coupling to the two degenerate
localized zero-energy states $\Sigma_0$ shown in Fig.~7(a,b) and
7(e,f). The interacting Hamiltonian $H_{\Sigma_0}$ is expressed as

\begin{small}
\begin{eqnarray}
& &H_{\Sigma_0}\\ \nonumber &= &\sum_{j,\sigma}E_{j}
c^{\dagger}_{j,\sigma}c_{j,\sigma}
+\sum_{j}U_j~n_{j,\sigma}n_{j,-\sigma}\\ \nonumber
&+&\frac{1}{2}\sum_{j\neq\ell,\sigma,\sigma'}U_{j,\ell}~
n_{j,\sigma}n_{\ell,\sigma'},
\end{eqnarray}
\end{small}

where $E_j=E_0=0$ represents the spin-independent energy level of
the $j$th zero-energy orbital. The intra-level and inter-level
Coulomb interactions ($U_j=U_0$ and $U_{j,\ell}=U_1$) are
evaluated by using the expression
\begin{eqnarray}
    &&U_{\ell=0,1}\\ \nonumber
    &=&\frac{1}{4\pi
        \kappa}\sum_{i,j}\frac{|\Psi^0_{1(2)}(\textbf{r}_i)|^2
        |\Psi^0_{1(2)}(\textbf{r}_j)|^2}{|\textbf{r}_i-\textbf{r}_j|},
\end{eqnarray}
where $\kappa = 4$ is the dielectric constant and $U_{cc} = 4$~eV
is the on-site Coulomb interaction corresponding to double
occupancy in each $p_z$ orbital. In Eq. (4), the wave functions
are
$\Psi^0_{1}=\frac{\varphi_{B1}+\varphi_{B2}-2\varphi_{B3}}{\sqrt{6}}$
and $\Psi^0_{2} =\frac{\varphi_{B1}-\varphi_{B2}}{\sqrt{2}}$. For
simplicity, we employ the BAT molecule to illustrate the
experimental tunneling current spectra. We obtain $U_0=0.25$~eV
and $U_{1}=0.08$~eV for the state $\Psi^0_{1}$. The value $U_0 =
0.37$~eV for $\Psi^0_{2}$ is considerably larger than that for
$\Psi^0_{1}$. In calculating the tunneling current spectra, we
neglect this difference in the intra-level Coulomb interactions
and use $U_0 = 0.25$~eV, since the applied voltage $V_a$ in Ref.
[\onlinecite{Pascual}] satisfies $|V_{a}| \le 200 ~mV$. The number
operator is defined as
$n_{j,\sigma}=c^\dagger_{j,\sigma}c_{j,\sigma}$. Because we focus
on the low-temperature transport regime, thermal excitation
processes involving higher-energy states can be neglected.

The nonequilibrium Green-function method provides an effective
framework for calculating tunneling currents through correlated
nanostructures under finite bias
[\onlinecite{DavidK1}--\onlinecite{Kuo6}]. Using the
equation-of-motion approach, the tunneling current through the
$\Sigma_0$ states in the presence of electron-electron
interactions is given by

\begin{eqnarray}
& &J_{STM}(V_a)\\ \nonumber &=&\frac{2e}{h}\int {d\varepsilon}~
{\cal T}_{STM}(\varepsilon) [f_S(\varepsilon)-f_D(\varepsilon)],
\end{eqnarray}
where the source electrode corresponds to the STM tip and the
drain electrode represents the metallic substrate. The Fermi-Dirac
distribution of electrode $\alpha$ is $ f_{\alpha}(\varepsilon)=
\frac{1}{\exp[(\varepsilon-\mu_{\alpha})/k_BT]+1}$, with chemical
potentials $\mu_{S(D)}=\mu\pm eV_a/2$ under an applied bias of
$+V_a/2$ and $-V_a/2$. The analytical expression for ${\cal
T}_{STM}(\varepsilon)$ is provided in Appendix~C. The present
treatment focuses on the molecular low-energy states and Coulomb
blockade transport; an explicit treatment of Kondo correlations in
the molecule-substrate junction in beyond scope of the present
work[\onlinecite{MadhavanV}--\onlinecite{BaoZQ}].

Figures~8(a)-8(c) show the total occupation number ($N_t$), the
tunneling current ($J$), and the differential conductance
($dJ/dV_a$) as functions of the applied voltage $V_a$ for
different drain coupling strengths $\Gamma_D$. The calculations
are performed with $\Gamma_S=\gamma=0.1$~meV, temperature
$T=12$~K, and $\mu=E_0=0$. Here, $\Gamma_S$ and $\Gamma_D$
represent the tunneling rates between the STM tip (source) and
metallic substrate (drain), respectively, and the two degenerate
$\Sigma_0$ energy levels.

As shown in Fig.~8(a), the total occupation number exhibits a
strong bias-direction dependence when $\Gamma_S\neq\Gamma_D$. For
$\Gamma_S=\gamma$ and $\Gamma_D=0.1\gamma$, electron occupation is
strongly enhanced under backward bias, corresponding to the
shell-filling regime. In contrast, under forward bias the
occupation of the $\Sigma_0$ states is suppressed because
electrons tunnel through the molecular states without significant
accumulation, corresponding to the shell-tunneling regime.

Figure~8(b) presents the corresponding tunneling current. The
largest current is obtained for symmetric coupling
$\Gamma_S=\Gamma_D$, where the current-voltage characteristics are
symmetric with respect to the bias direction. When
$\Gamma_S\neq\Gamma_D$, the current becomes dominated by the
shell-tunneling process. The current plateau observed within
$|V_a|\leq160$~mV originates from the inter-level Coulomb
interaction $U_{\ell,j}$ between the two zero-energy states.

The differential conductance spectra shown in Fig.~8(c) exhibit
three characteristic peaks. The central peak $\varepsilon_0$
corresponds to the zero-bias differential conductance associated
with tunneling through the degenerate $\Sigma_0$ states. When
$\Gamma_S \neq \Gamma_D$, this peak becomes weaker and deviates
from the Lorentzian line shape. Similar asymmetric
differential-conductance spectra have also been observed
experimentally in Figure S3 of Ref. [\onlinecite{Pascual}]. In
addition, no negative differential conductance is observed in our
calculations, consistent with the experimental measurements. For
$\Gamma_S=\gamma$ and $\Gamma_D=0.1\gamma$, the differential
conductance under backward bias ($\varepsilon_{1,B}$) is
significantly larger than that under forward bias
($\varepsilon_{1,F}$). Both features originate from charge
tunneling through the excited energy level $E_0+U_{\ell,j}$
induced by inter-level Coulomb interactions.

Comparison with the experimental differential-conductance spectra
of AAT and BAT molecules further reveals the important role of the
substrate environment. The nearly symmetric
differential-conductance spectrum of AAT reported in
Ref.~[\onlinecite{Pascual}] corresponds to the symmetric coupling
condition $\Gamma_S=\Gamma_D$, whereas the strongly asymmetric BAT
spectrum is consistent with the asymmetric coupling regime
$\Gamma_S\gg\Gamma_D$. Therefore, our results demonstrate that the
STM tunneling characteristics of AAT and BAT molecules are not
determined solely by their intrinsic molecular orbitals but are
also strongly influenced by the coupling conditions between the
molecule, STM tip, and metallic substrate. The experimental
differential electrical conductance spectra ($dJ/dV_a$) exhibit a
sharp, narrow peak near zero bias, whereas the peaks in the
high-bias region are considerably broader. These features may be
associated in part with bias-dependent tunneling rates. The
experimental conductance spectra also depend on the positions of
the STM tip contacted zigzag-edge sites. These factors limit the
extent to which our present model can quantitatively reproduce the
experimental observables. Nevertheless, the comparison provides a
qualitative assessment of the consistency between the calculated
tunneling spectra and the experimental observations.

\begin{figure}[h]
\centering
\includegraphics[trim=1.cm 0cm 1.cm 0cm,clip,angle=0,scale=0.25]{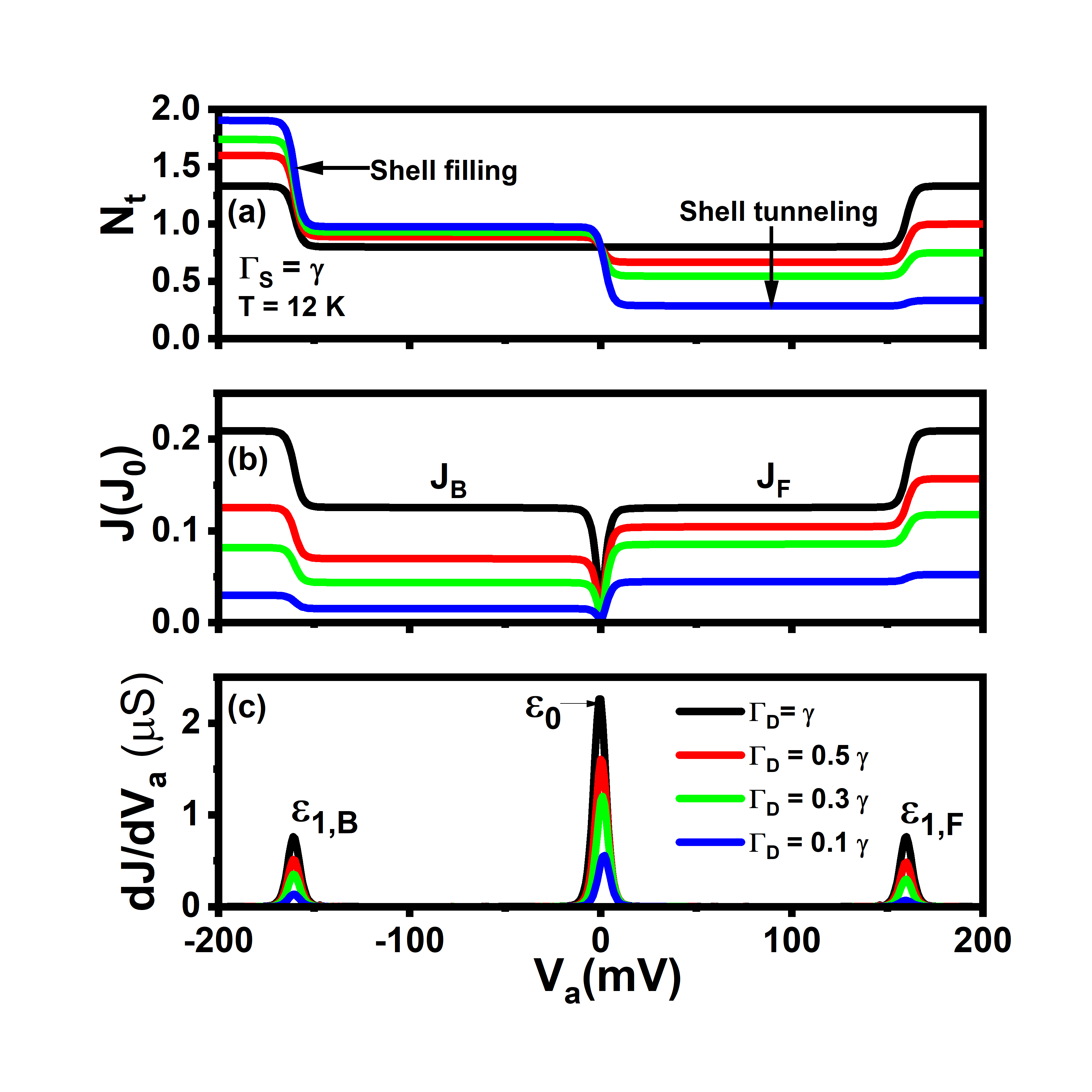}
\caption{(a) Total occupation number,
$N_t=\sum_{\sigma}(N_{1,\sigma}+N_{2,\sigma})$, (b) tunneling
current, $J_{STM} = J$, and (c) differential conductance,
$dJ/dV_a$, as functions of the applied voltage $V_a$ for various
tunneling rates $\Gamma_{1,D}=\Gamma_{2,D}=\Gamma_D$ at
$\Gamma_{1,S}=\Gamma_{2,S}=\Gamma_S=\gamma=0.1$~meV and $T=12$~K.
The tunneling current and differential conductance are expressed
in units of  $J_0=0.773$~nA and $\mu$S, respectively.}
\end{figure}

\subsection{Artificial quantum materials}

Artificial quantum materials (AQMs) have recently attracted
significant attention because they provide a versatile platform
for engineering quantum phenomena beyond those available in
naturally occurring materials. Using bottom-up molecular synthesis
techniques, researchers have pursued artificial structures
exhibiting exotic quantum phases, including fractional quantum
Hall states, superconductivity, and magnetic ordering
[\onlinecite{HenriquesJ}--\onlinecite{YanY},\onlinecite{NaritsukaM}--\onlinecite{ShenLG}].
In parallel, the design of artificial flat-band systems has become
an active area of research because flat bands provide a natural
platform for realizing compact localized states (CLSs) induced by
destructive quantum interference and exploring their potential
applications in quantum information storage
[\onlinecite{RomerRA}]. Although flat bands have been
experimentally demonstrated in several engineered systems, most
existing realizations involve bosonic particles
[\onlinecite{JoGB}--\onlinecite{LeykamD}], while electronic
flat-band materials remain challenging to achieve
[\onlinecite{YanY},\onlinecite{LeykamD}--\onlinecite{YinRT}].

Motivated by the experimentally synthesized triangulene-based
nanographenes in Ref.~[\onlinecite{Pascual}] and the protected
junction orbitals identified in the $3$T-AGNR/$7$-AGNR/$3$T-AGNR
structures, we propose artificial zigzag graphene nanoribbon
(aZGNR) and artificial armchair graphene nanoribbon (aAGNR)
segments, as illustrated in Fig.~9. The junction orbitals in these
artificial structures are protected by the chiral symmetry of
graphene and remain well separated from the bulk continuum states,
providing robust localized building blocks for constructing AQMs.
Since the effective bond lengths between neighboring orbitals can
be precisely tuned, the corresponding electron hopping energies in
aZGNRs and aAGNRs can be reduced to the meV scale. In particular,
the engineered flat subbands in aAGNRs emerge close to the Fermi
energy, offering an electronic platform for realizing tunable CLSs
and exploring strongly correlated quantum phases.

\begin{figure}[h]
\centering
\includegraphics[trim=1.cm 0cm 1.cm 0cm,clip,angle=0,scale=0.25]{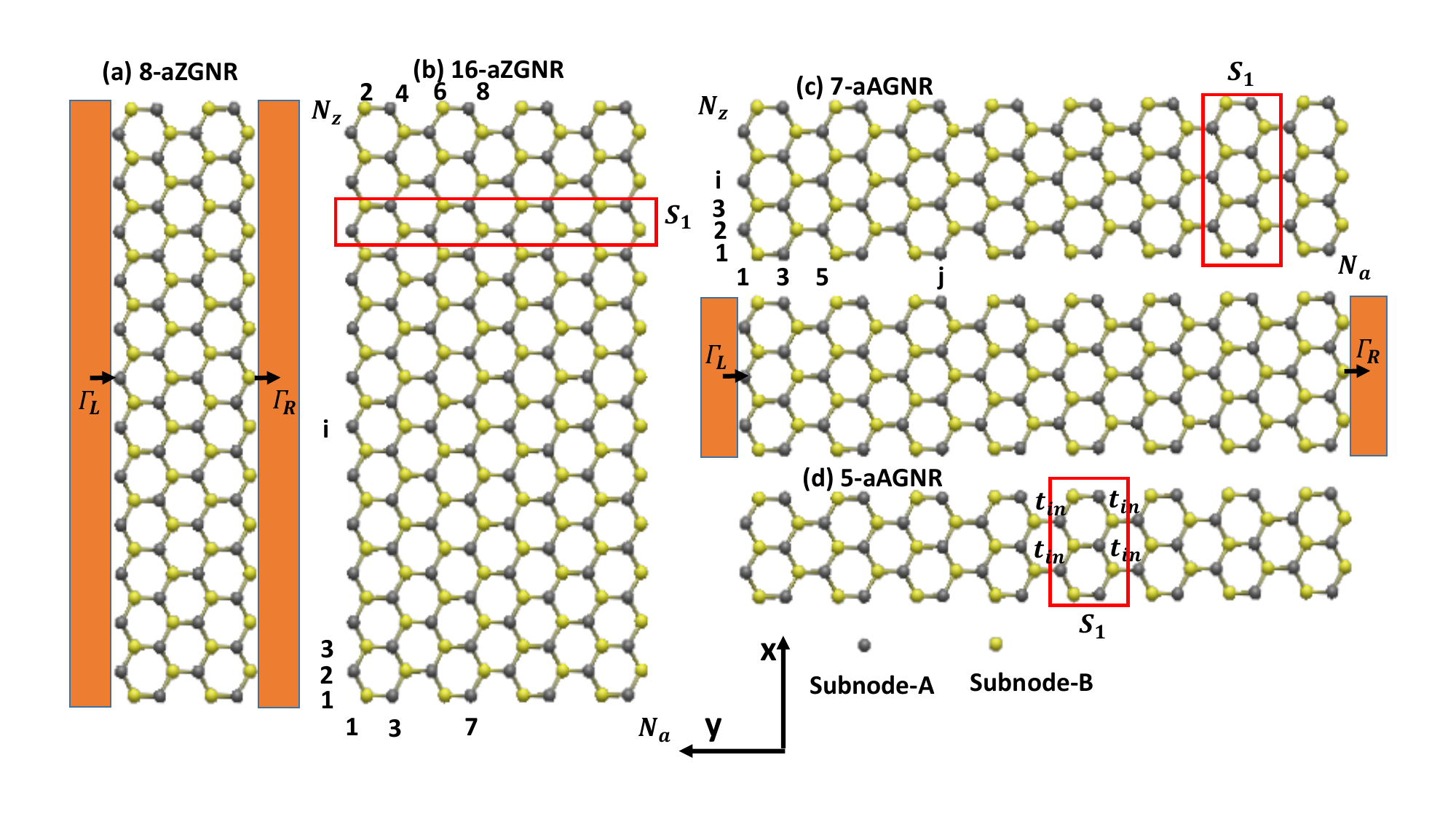}
\caption{ Geometry of artificial quantum materials (AQMs). (a) and
(b) Schematic illustrations of artificial zigzag graphene
nanoribbons (aZGNRs). (c) and (d) Schematic illustrations of
artificial armchair graphene nanoribbons (aAGNRs). These AQMs are
characterized by $(N_z,N_a)$, where $N_z$ and $N_a$ denote the
numbers of localized node orbitals along the zigzag and armchair
edge directions, respectively. $S_1$ represents the unit cell of
the corresponding aZGNR and aAGNR structures. $\Gamma_{L(R)}$
denote the tunneling rates for electrons tunneling from the left
(right) electrode into the adjacent nodes at the zigzag edges.
Here, the elementary building block is chosen as the one-orbital
node realized by the 3T-7-AGNR/7-AGNR structure, which can be
experimentally synthesized and chemically tuned through
bond-length engineering. Unlike pristine graphene with a fixed
carbon-carbon distance $a_{cc}=1.42~\AA$, the effective lattice
constants of aZGNRs and aAGNRs can be continuously tuned. For the
$y=4~R_1$ configuration, the distance between neighboring nodes is
$12a_{cc}=1.7$~nm, giving an effective hopping parameter of
$t_{SL}=30$~meV. These parameters are used throughout this
section.}
\end{figure}

\textbf{1. Artificial ZGNR structures}

The electronic structures of conventional graphene nanoribbons
described by the $p_z$ one-orbital tight-binding model are
reliable mainly near the conduction-band minimum (CBM) and
valence-band maximum (VBM), because electronic states far away
from these band edges can be significantly modified by the
contribution of $\sigma$ bands
[\onlinecite{SevincliH}--\onlinecite{LinHH}]. In particular, the
flat bands predicted in conventional AGNRs are located far away
from the CBM and VBM [\onlinecite{ZhengHX},\onlinecite{LinHH}],
making it difficult to experimentally explore quantum interference
effects and many-body phenomena associated with these flat-band
states [\onlinecite{AlmeidaPA}].

In contrast, the AQMs proposed here are constructed from
one-orbital nodes derived from the low-energy interface states of
$n$T-AGNR/AGNR structures. Therefore, the resulting artificial
subbands originate directly from the low-energy electronic modes
near the CNP and provide a more realistic platform for
investigating flat-band physics. The bandwidth of these artificial
subbands can be continuously controlled by tuning the effective
distance between neighboring nodes, which determines the hopping
parameter $t_{SL}$ (see Appendix~B).

Figure~10 presents the electronic structures of aZGNRs with
different node widths characterized by $N_a$. Similar to pristine
ZGNRs, aZGNRs remain metallic, while the maximum bandwidth of the
artificial subbands is determined by the effective hopping
strength and is approximately $\pm~3t_{SL}$. The first conduction
and valence subbands within the momentum range $\frac{2\pi}{3}\leq
|k_x|\leq\pi$ originate from localized zigzag edge orbitals and
are responsible for the magnetic edge states observed in
conventional ZGNRs
[\onlinecite{NakadaK}--\onlinecite{WakabayashiK2}]. The second
conduction and valence subbands exhibit valley degeneracy. The
corresponding saddle points in the energy dispersion generate
van-Hove singularities in the density of states
(DOS)[\onlinecite{HodO}]. Such valley degenerate electronic
structures may provide potential applications in valleytronics and
thermoelectric devices [\onlinecite{SuiM}].

\begin{figure}[h]
\centering
\includegraphics[trim=1.cm 0cm 1.cm 0cm,clip,angle=0,scale=0.3]{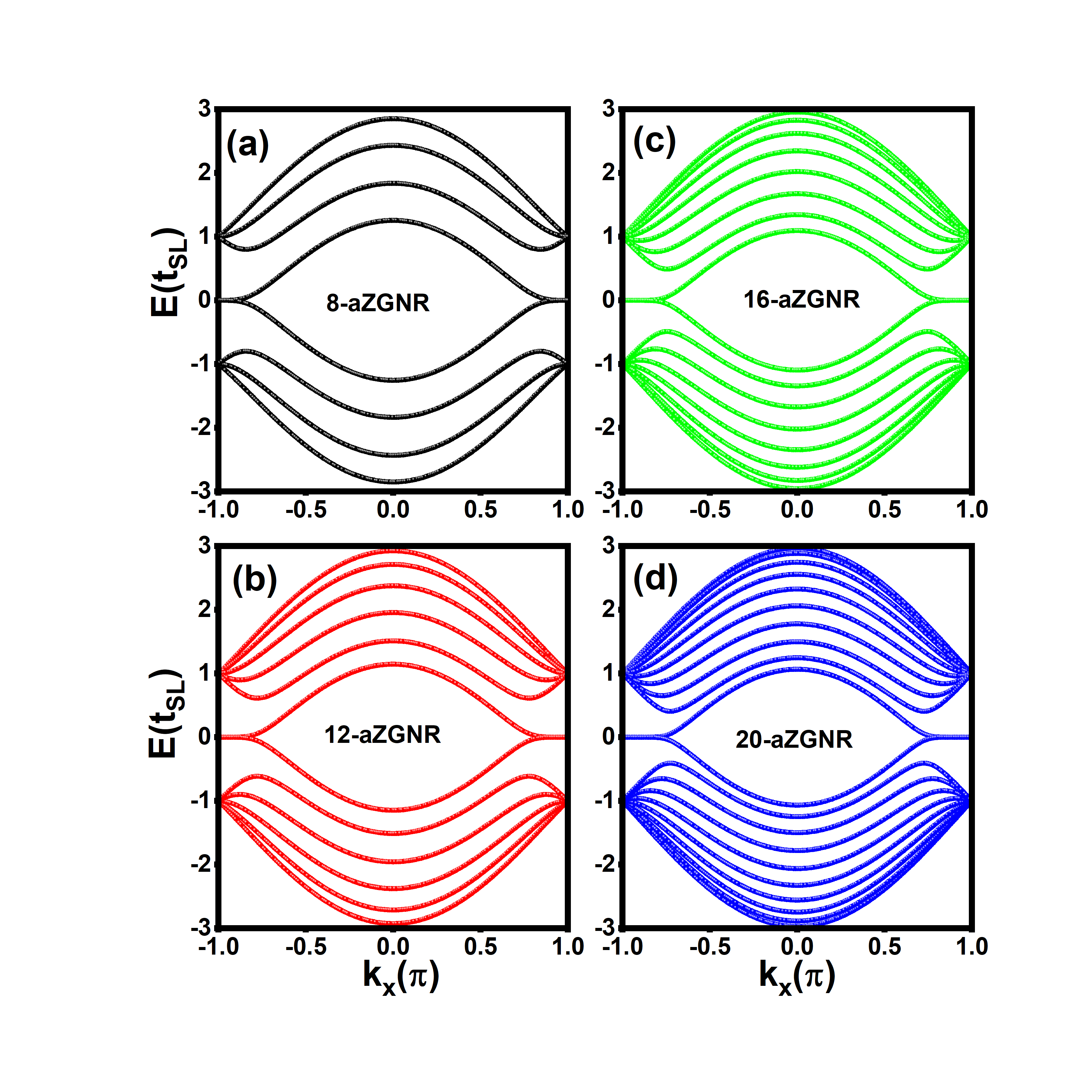}
\caption{Band structures of artificial zigzag graphene nanoribbons
(aZGNRs): (a) 8-aZGNR, (b) 12-aZGNR, (c) 16-aZGNR, and (d)
20-aZGNR. The unit cell length $S_1$ is set to one for
simplicity.}
\end{figure}

Because large-scale structures remain challenging to fabricate
using bottom-up synthesis techniques, experimentally accessible
AQMs are expected to have finite dimensions, similar to the
nanographene structures synthesized in
Ref.~[\onlinecite{Pascual}]. To demonstrate the observable
signatures of van-Hove singularities in finite aZGNR systems, we
calculate the transmission coefficient ${\cal
T}_{Zig}(\varepsilon)$ of finite 8-aZGNR segments connected to
electrodes through their zigzag edges, as shown in Fig.~11.
Figures~11(a)-11(d) correspond to aZGNR lengths characterized by
$N_z=65$, $79$, $93$, and $107$, respectively. For $N_z=107$, the
length of the aZGNR segment exceeds 100~nm. Six pronounced
transmission peaks ($\varepsilon\geq0$), labeled as
$\rho_{i=1,\ldots,6}$, are observed and correspond to the six
saddle points in the electronic dispersion. Since the transmission
coefficient ${\cal T}_{Zig}(\varepsilon)$ is directly related to
the electrical conductance at zero temperature, these results
indicate that transport measurements of finite aZGNR segments can
provide an experimental route for detecting the predicted van-Hove
singularities [\onlinecite{HodO}].

\begin{figure}[h]
\centering
\includegraphics[angle=0,scale=0.3]{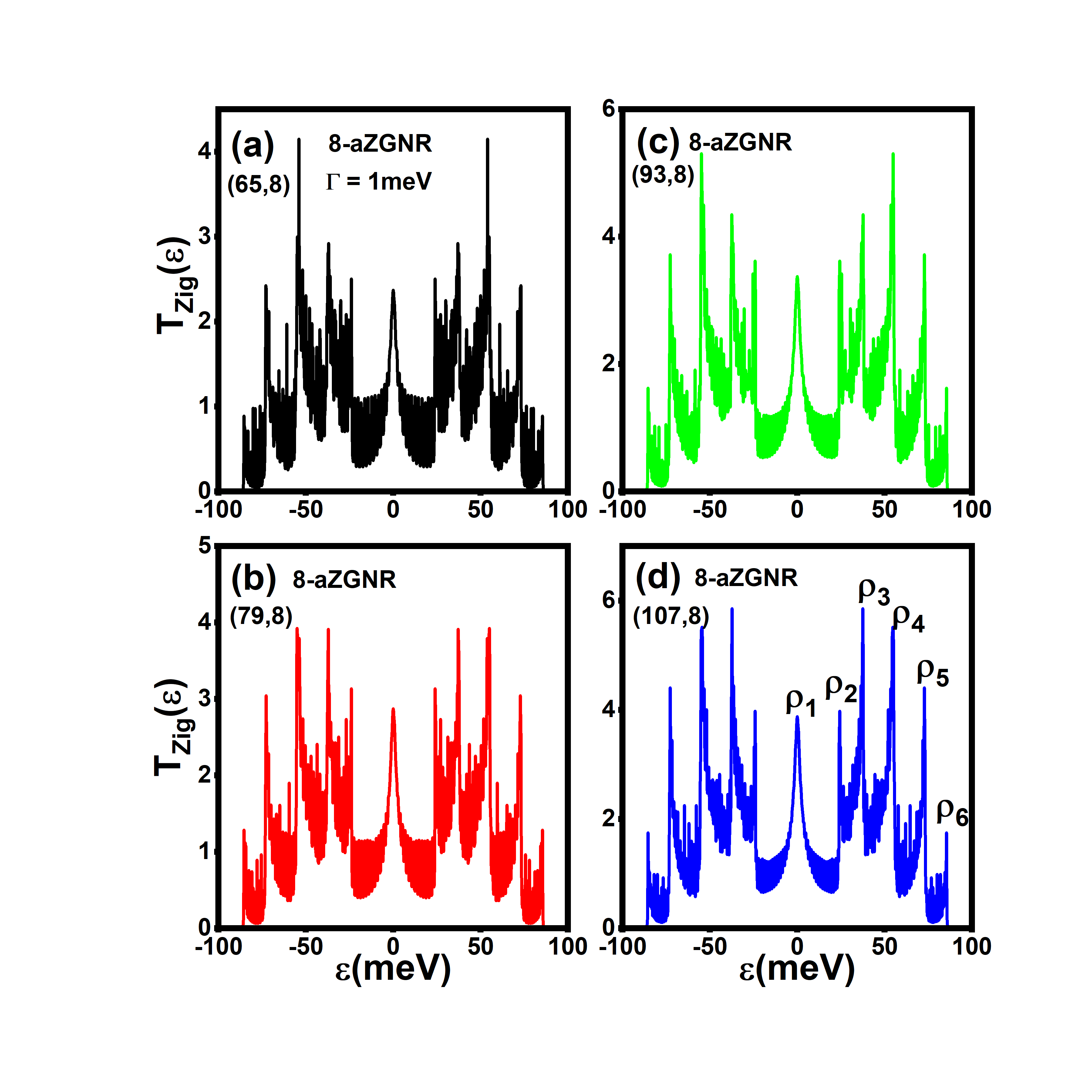}
\caption{Transmission coefficient ${\cal T}_{Zig}(\varepsilon)$ of
the 8-aZGNR segment for different values of $N_z$ at
$\Gamma_L=\Gamma_R=\Gamma=1$~meV: (a) $N_z = 65$, (b) $N_z = 79$,
(c) $N_z = 93$, and (d) $N_z = 107$. Here, $t_{SL} = 30$~meV, and
the aZGNR is coupled to the electrodes through its zigzag edges.}
\end{figure}

\textbf{2. Artificial AGNR structures}

Although aZGNR segments possess localized zigzag-edge states,
their electronic dispersions remain strongly dispersive. In
contrast, aAGNRs provide a platform for realizing nearly
dispersionless subbands through the interference of artificial
orbitals. Figure~12 shows the calculated electronic structures of
aAGNRs with different node widths ($N_z$). Similar to conventional
AGNRs, the electronic phases of aAGNRs are determined by their
width [\onlinecite{NakadaK}--\onlinecite{WakabayashiK2}]. As shown
in Figs.~12(a)-12(d), 5-aAGNR and 11-aAGNR exhibit metallic
behavior, whereas 7-aAGNR and 9-aAGNR are semiconducting.

For odd-node-width aAGNRs, artificial flat subbands emerge at
$E=\pm t_{SL}$ [\onlinecite{ZhengHX}]. Since the effective hopping
parameter $t_{SL}$ between neighboring artificial orbitals is in
the meV range, these flat subbands are located very close to the
CNP (or Fermi energy). Therefore, unlike the conventional AGNR
flat bands that appear far away from the low-energy region, the
artificial flat subbands proposed here provide a realistic
platform for investigating quantum interference and many-body
effects[\onlinecite{LinHH}]. To the best of our knowledge, charge
transport through such engineered electronic flat subbands has not
been theoretically or experimentally explored, although related
electronic phases have been investigated in various designed
nanographene systems
[\onlinecite{AlmeidaPA},\onlinecite{WangJN},\onlinecite{YuanXY}].

\begin{figure}[h]
\centering
\includegraphics[trim=1.cm 0cm 1.cm 0cm,clip,angle=0,scale=0.3]{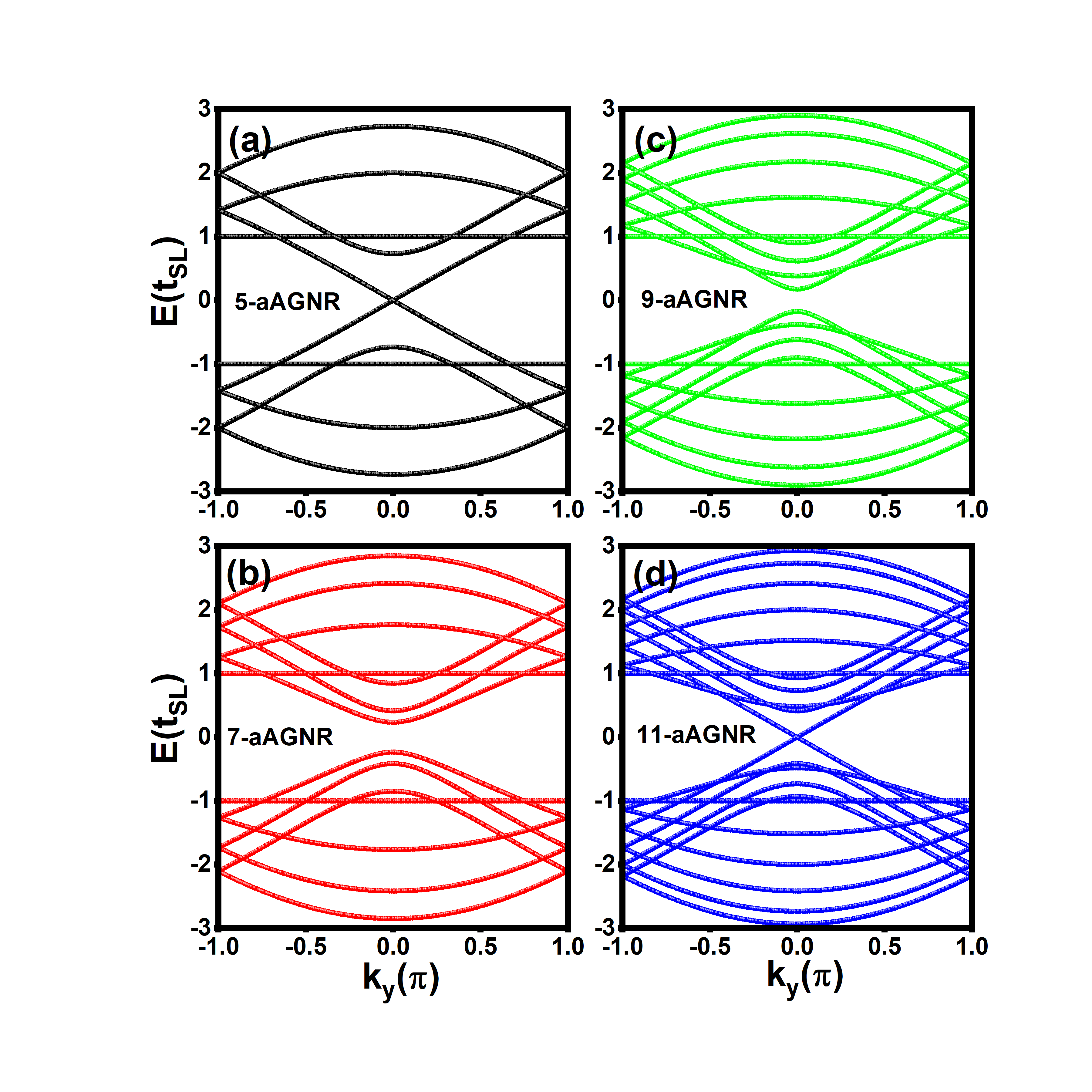}
\caption{Band structures of artificial armchair graphene
nanoribbons (aAGNRs): (a) 5-aAGNR, (b) 7-aAGNR, (c) 9-aAGNR, and
(d) 11-aAGNR. The unit cell length $S_1$ is set to one for
simplicity.}
\end{figure}

To explore the transport signatures of these flat subbands, we
calculate the transmission coefficient ${\cal
T}_{Zig}(\varepsilon)$ of finite 7-aAGNR segments with zigzag-edge
contacts to electrodes. The results for different lengths
characterized by $N_a$ are presented in Fig.~13. To clearly
resolve the flat-subband contribution, we focus on the energy
range near $\varepsilon=t_{SL}$. Due to the finite size of the
system, the energy spectrum becomes quantized. For $N_a=60$
[Fig.~13(a)], the transmission coefficients are strongly
suppressed because of backward-scattering components. However, a
pronounced transmission feature unexpectedly appears at the
flat-subband energy $\varepsilon=t_{SL}=30$~meV. For $N_a=64$ and
$N_a=72$, the transmission at $\varepsilon = t_{SL}$ becomes
almost completely suppressed [Figs.~13(b) and 13(d)]. In contrast,
significant transmission is recovered for $N_a = 68$ [Fig.~13(c)].
At the energies $\varepsilon=\pm t_{SL}$, two types of states
coexist: localized states associated with the flat subbands and
extended bulk states belonging to other subbands. Since the group
velocity of the dispersionless flat subbands approaches zero,
their contribution to the transmission is negligible. Therefore,
the pronounced transmission peaks observed at $\varepsilon=t_{SL}$
should originate from the extended bulk states of the other
subbands.

\begin{figure}[h]
\centering
\includegraphics[angle=0,scale=0.3]{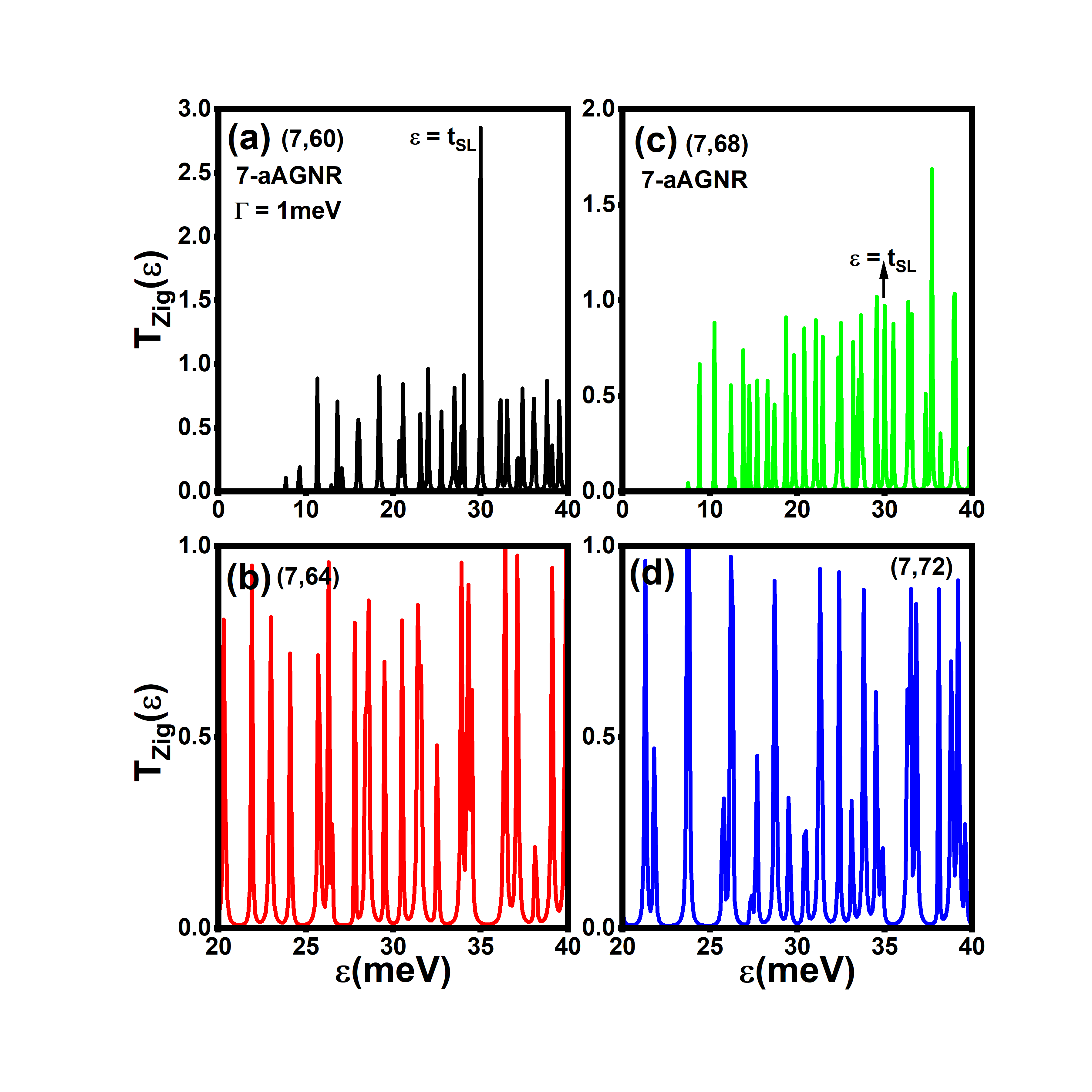}
\caption{ Transmission coefficient ${\cal T}_{Zig}(\varepsilon)$
of the 7-aAGNR segment for different values of $N_a$ at
$\Gamma_L=\Gamma_R=\Gamma=1$~meV: (a) $N_a=60$, (b) $N_a=64$, (c)
$N_a=68$, and (d) $N_a=72$. Here, $t_{SL}=30$~meV, and the aAGNR
is coupled to the electrodes through its zigzag edges.}
\end{figure}

To further clarify the transport properties of the flat subbands,
we calculate the transmission coefficient ${\cal
T}_{Arm}(\varepsilon)$ of 7-aAGNR segments with armchair-edge
contacts, which probes electron transport along the zigzag-node
direction. Figures~14(a)-14(d) correspond directly to the
structures shown in Fig.~13(a)-13(d). In Fig.~14, pronounced
transmission peaks $\sigma_C$ and $\sigma_V$ appear at
$\varepsilon=\pm t_{SL}=30$~meV, corresponding to the engineered
flat subbands. The magnitude of these peaks is determined by the
number of unit cells in the aAGNR segment, $N_s=N_a/4$. In
addition, the feature labeled as $\Sigma_0=2$ originates from two
additional zero-energy end states of the 7-aAGNR segment. The
spectra also exhibit van-Hove-type DOS features labeled as
$\rho_{VH}$.

A detailed analysis reveals a unique length dependence of the
flat-subband transport. For $N_a=64$ and $N_a=72$, corresponding
to $N_s=16$ and $N_s=18$, respectively, the peak amplitudes
satisfy $\sigma_{C(V)}=N_s$. In contrast, for $N_a=60$ and
$N_a=68$, corresponding to odd values of $N_s$, the peak
amplitudes become $\sigma_{C(V)}=N_s+1$. The enhanced transmission
for odd $N_s$ indicates that the bulk states at $\varepsilon=\pm
t_{SL}$ provide conducting channels not only along the armchair
direction but also through the zigzag-edge direction. Additional
calculations presented in Appendix~D further demonstrate that the
CLSs associated with the flat-subband states remain robust under
different electrode-contact configurations.

\begin{figure}[h]
\centering
\includegraphics[angle=0,scale=0.30]{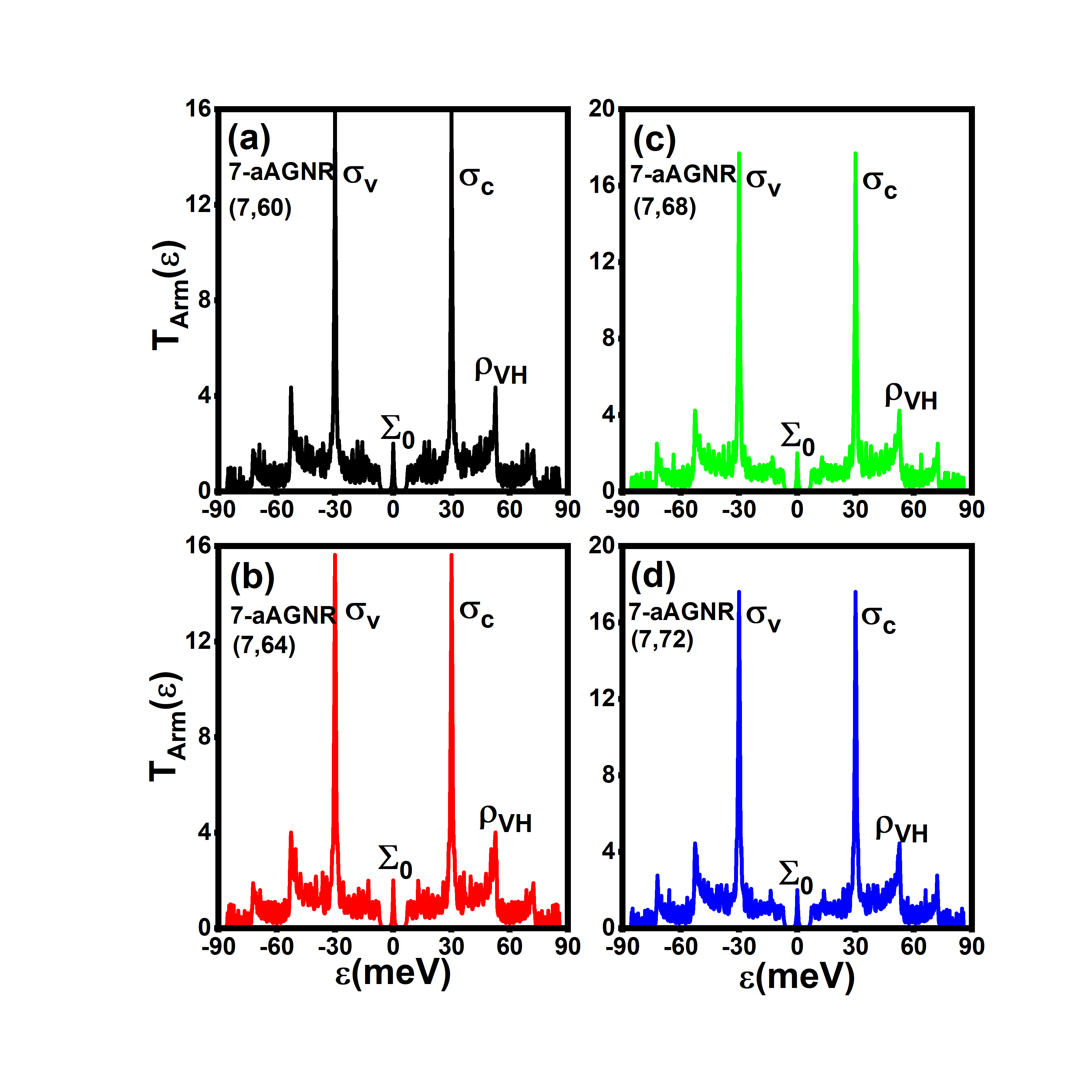}
\caption{Transmission coefficient, ${\cal T}_{Arm}(\varepsilon)$,
of 7-aAGNR segments with different values of $N_a$, at
$\Gamma_L=\Gamma_R= \Gamma = 1$ meV and $t_{SL} = 30$ meV: (a)
$N_a = 60$, (b) $N_a = 64$, (c) $N_a = 68$, and (d) $N_a = 72$.
The armchair edges of the aAGNR segments are connected to the left
and right electrodes. $\Gamma_{L(R)}$ denotes the tunneling rate
between the left (right) electrode and the adjacent nodes at the
armchair-terminated edge.}
\end{figure}

To illustrate the results of Fig. 13 and Fig. 14, we calculate the
charge density distributions of the energy levels at $\varepsilon
= t_{SL}$ for $N_a=60$, $64$, $68$, and $72$, as shown in Fig.~15.
For even $N_s$ values [Figs.~15(b) and 15(d)], the charge
densities are localized only on odd rows, $\ell = 1,3,5,7$, and
can be reproduced by repeating the charge distribution of a single
unit cell. Within each $S_1$ unit cell, the bond-pair charge
densities located on the subnode-A and subnode-B sites form CLSs
with energy $E=\pm t_{SL}$ induced by destructive quantum
interference (see Fig. 16). Consequently, the number of degenerate
flat-band states is equal to the number of unit cells, $N_s$.

In contrast, for odd $N_s$ values [Figs.~15(a) and 15(c)], the
charge-density distributions cannot be reproduced by a single unit
cell. For $N_s = 15$ and $17$, additional charge densities appear
at the zigzag-edge sites of specific unit cells. Specifically, for
$N_s = 15$, finite charge densities emerge on rows $\ell = 2, 4,
6$, whereas for $N_s = 17$, they are found only on rows $\ell = 2$
and $6$. These additional charge-density components originate from
bulk states belonging to other subbands. To examine whether this
behavior is unique to 7-aAGNRs, we further investigate aAGNR
segments with other odd node widths, including 3-, 5-, 9-, 11-,
and 13-aAGNRs. The 3-aAGNR is the narrowest aAGNR segment in which
the charge-density distributions at $\varepsilon = \pm t_{SL}$
receive contributions from both flat-subband states and bulk
states of other subbands (see Appendix~E). In the strong
interacting limit ($U_0 \gg t_{SL}$), the emergence of
ferromagnetic or Curie-type paramagnetic phases in aAGNR segments
depends sensitively on the number of unit cells, $N_s$
[\onlinecite{LinHH}]. The ferromagnetic states associated with
flat-subbands at $E = \pm t_{SL}$ depend on whether $N_s$ is even
or odd. We therefore propose exploiting this parity-controlled
magnetic switching mechanism for potential applications in quantum
memory devices[\onlinecite{RomerRA}].

\begin{figure}[h]
\centering
\includegraphics[angle=0,scale=0.3]{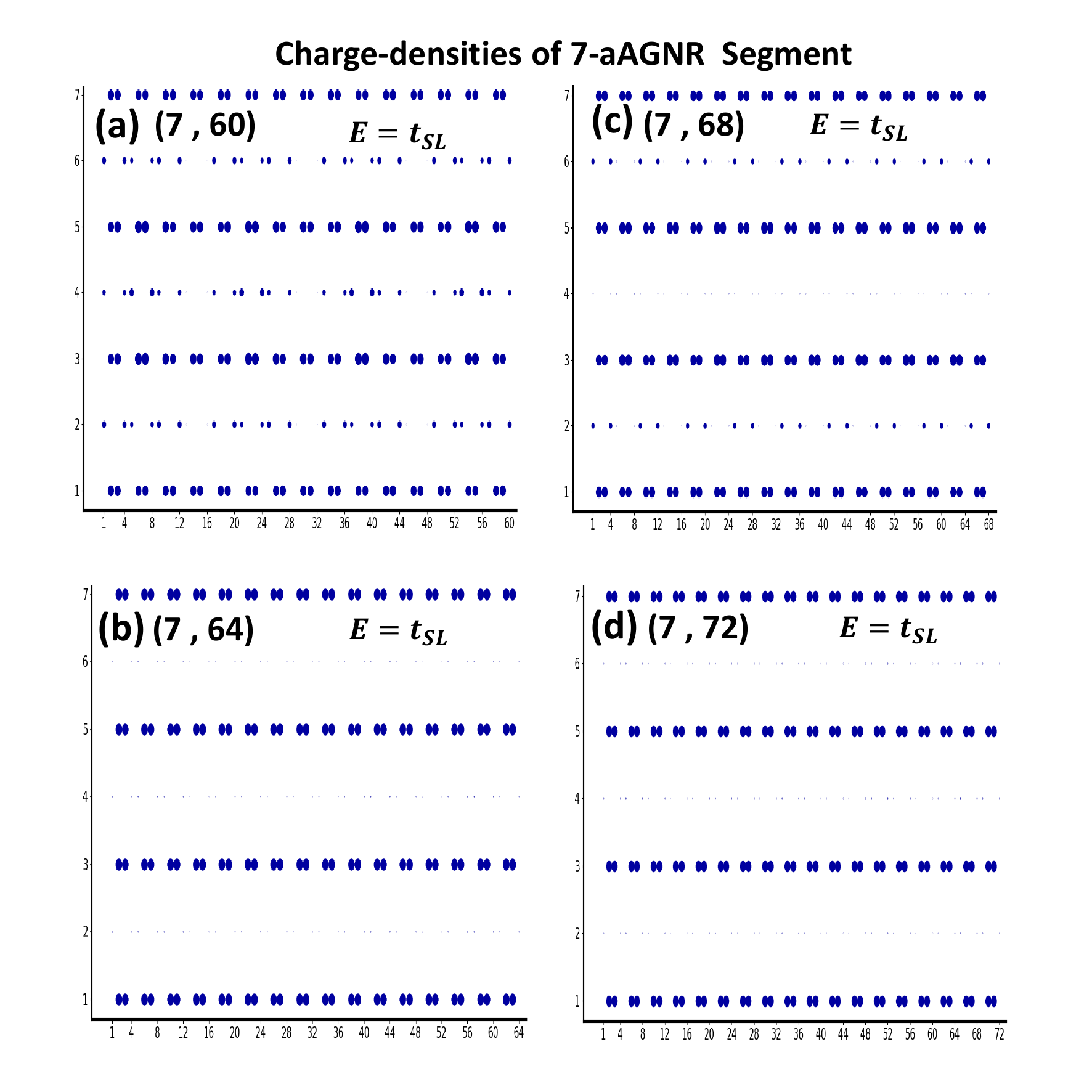}
\caption{Charge density distributions at $E = t_{SL}$ for 7-aAGNR
segments with different ribbon lengths: (a) $N_a = 60$ ($N_s =
15$), (b) $N_a = 64$ ($N_s = 16$), (c) $N_a = 68$ ($N_s = 17$),
and (d) $N_a = 72$ ($N_s = 18$). The size of each blue marker
represents the local spectral weight at the corresponding lattice
site at $E=t_{SL}$. The local spectral density is calculated as
$n(\textbf{x})=\frac{-1}{\pi}\mathrm{Im}G^r_{j,\ell}(\varepsilon=t_{SL})$,
with $\Gamma_{L(R)}\rightarrow 0$. }
\end{figure}

\begin{figure}[h]
\centering
\includegraphics[angle=0,scale=0.3]{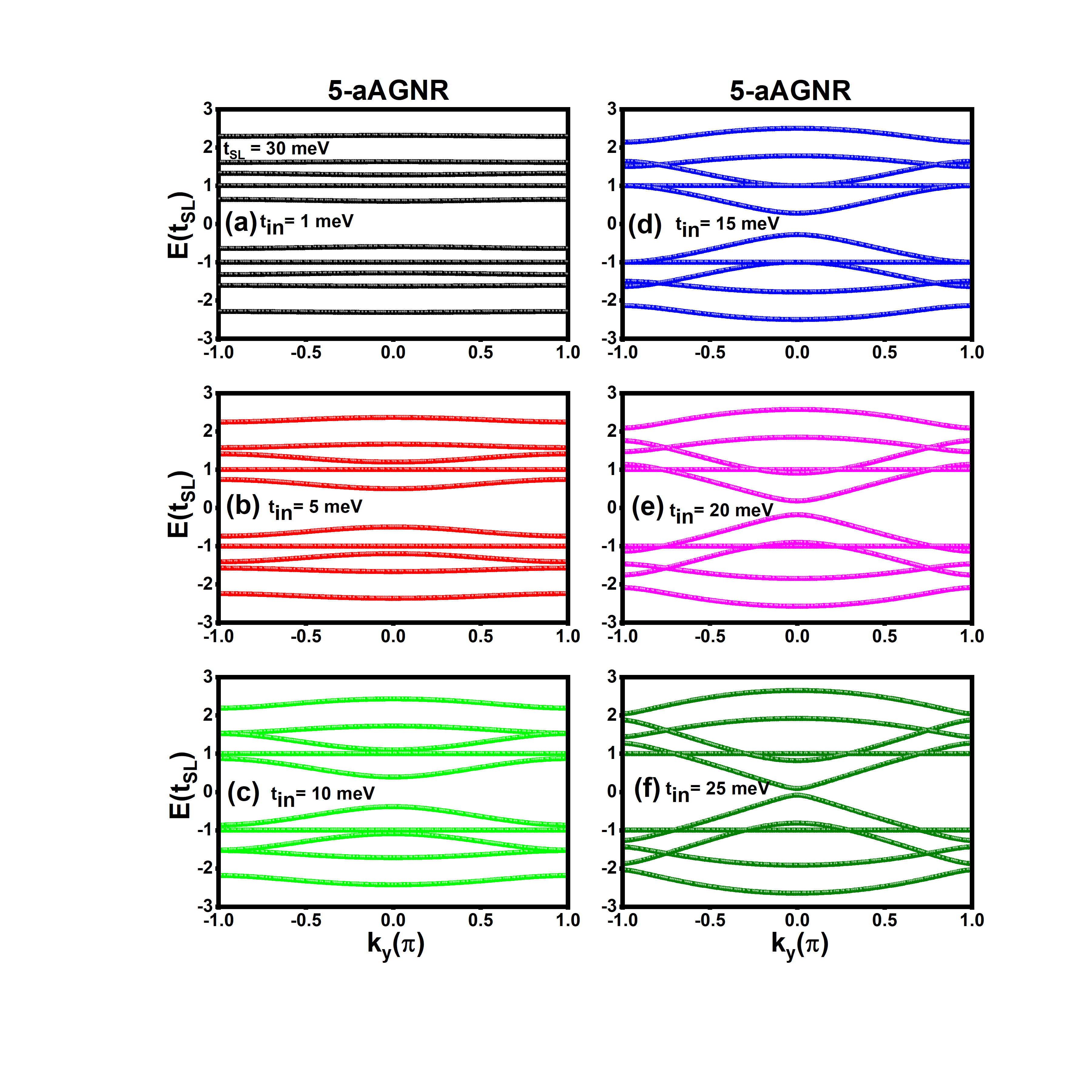}
\caption{Electronic band structures of the 5-aAGNR with
$t_{SL}=30$~meV for different inter-unit-cell hopping strengths:
(a) $t_{in}=1$~meV, (b) $t_{in}=5$~meV, (c) $t_{in}=10$~meV, (d)
$t_{in}=15$~meV, (e) $t_{in}=20$~meV, and (f) $t_{in}=25$~meV.}
\end{figure}

\textbf{3. Compact localized states induced by destructive quantum
interference}

As illustrated in Fig.~15, the charge density associated with the
flat subbands vanishes at the node sites connecting neighboring
$S_1$ unit cells. This characteristic indicates that the wave
functions of these CLSs are confined within individual unit cells
and do not extend through the inter-unit-cell
nodes[\onlinecite{Bergman}--\onlinecite{Morfonios}]. Consequently,
the flat subbands are expected to be insensitive to the
inter-unit-cell hopping strength $t_{in}$ (see Fig. 9(d)). As
$t_{in}$ is continuously increased from zero to $t_{SL}$, the flat
subbands remain pinned at $E = \pm t_{SL}$, whereas the dispersive
bulk subbands evolve significantly because their wave functions
extend across neighboring unit cells.

To verify this picture, Fig.~16 presents the electronic band
structures of the 5-aAGNR with $t_{SL} = 30$~meV for different
inter-unit-cell hopping strengths: (a) $t_{in} = 1$~meV, (b)
$t_{in} = 5$~meV, (c) $t_{in} = 10$~meV, (d) $t_{in} = 15$~meV,
(e) $t_{in} = 20$~meV, and (f) $t_{in} =2 5$~meV. For the
weak-coupling case of $t_{in}=1$~meV, ten narrow subbands are
observed, including a pair of nearly dispersionless flat subbands
located at $E = \pm t_{SL}$. These subbands are well separated
from the remaining narrow bands. As $t_{in}$ increases, the
dispersive narrow subbands broaden progressively owing to the
enhanced coupling between neighboring unit cells, whereas the flat
subbands remain essentially unchanged. At $t_{in} = 10$~meV, the
first conduction subband maximum (1-CSBM) and the third conduction
subband minimum (3-CSBM) become degenerate with the flat subband
at $E = t_{SL}$. Upon further increasing the coupling to $t_{in} =
20$~meV, the 1-CSBM moves above, while the 3-CSBM shifts below,
the flat-subband energy. Even for the strongest coupling
considered, $t_{in} = 25$~meV, the 5-aAGNR still retains a small
band gap.

The robustness of the flat-subband energies against variations in
$t_{in}$ provides compelling evidence that these flat subbands
originate from destructive quantum interference rather than weak
inter-unit-cell coupling. Because the wave-function amplitudes
vanish at the connecting nodes, electron hopping between
neighboring unit cells is effectively suppressed, giving rise to
CLSs protected by destructive interference.

At zero temperature, the electrical conductance is determined by
$G_e(\mu) = G_0{\cal T}(\mu)$, where $G_0=2e^2/h$ is the quantum
conductance and $\mu$ is the chemical potential. For the
flat-subband energies $\mu = \pm t_{SL}$, the conductance through
armchair-edge contacts is much larger than that through
zigzag-edge contacts, $G_{e,Arm}(\mu)\gg G_{e,Zig}(\mu)$. This
contact-direction-dependent transport originates from quantum
interference of CLSs and is analogous to interference effects
observed in molecular junctions
[\onlinecite{Guedon}--\onlinecite{Cardamone}].

\section{Conclusion}

In this work, we develop a real-space framework to uncover the
microscopic origin of localized interface states in graphene-based
artificial quantum materials composed of $n$-triangulenes (nT) and
armchair graphene nanoribbon (AGNR) segments. Unlike conventional
approaches relying on bulk topological invariants, our approach
directly reveals how triangulene zero-energy modes and AGNR end
states  continuously evolve into localized node orbitals through
tunable coupling between graphene building blocks. This real-space
picture provides a transparent understanding of compact localized
state (CLS) formation and establishes a general strategy for
engineering graphene-based artificial quantum materials.

For $n$-triangulenes, the number of zero-energy modes is governed
by the sublattice imbalance relation $(n-1)=|n_A-n_B|$, where
$n_A$ and $n_B$ represent the numbers of sublattice-A and
sublattice-B sites, respectively. The AGNR segments considered
here contain $R_1$-type and $R_2$-type unit cells, supporting one
and two end states per terminus, respectively. By continuously
tuning the junction hopping parameters, we demonstrate the
transformation of isolated graphene end states into
interface-localized orbitals in coupled three-arm graphene
junctions.

Using 3T-GNR/$7_y$-AGNR/3T-GNR and 4T-GNR/$7_y$-AGNR/4T-GNR
structures as representative examples, we establish a
general counting rule for determining the number and chirality of
node orbitals, $N_{node,\delta}=|N_{es,t,A(B)}-N_{tri,0,B(A)}|$,
where $N_{es,t}$ denotes the total number of AGNR end states
contributed by the three AGNR arms at the junction, and
$N_{tri,0}$ represents the number of zero-energy modes of the
attached triangulene. The chirality of the node orbitals is
governed by the dominant constituent: AGNR end states determine
the chirality when $N_{es,t}>N_{tri,0}$, whereas triangulene
zero-energy modes dominate when $N_{tri,0}>N_{es,t}$. Owing to
graphene sublattice symmetry, the chiralities of node orbitals at
the two opposite junctions remain reversed. Furthermore, the
different orbital compositions of 3T and 4T junctions provide
additional control over the localization characteristics of
artificial graphene structures.

The proposed framework also offers a natural interpretation of
experimentally synthesized triangulene-based nanographenes. For
AAT and BAT structures composed of 3T-$7_w$-AGNR/$7_y$-AGNR
junctions, the observed zero-energy modes are identified as
compact localized orbitals generated by destructive interference
among graphene end states. By combining an extended Anderson model
with the nonequilibrium Green's function formalism, we demonstrate
that the STM tunneling spectra of AAT and BAT molecules depend not
only on their intrinsic molecular orbitals but also on the
coupling conditions between the molecule, STM tip, and metallic
substrate. Moreover, these localized junction orbitals can support
long-distance coherent tunneling, suggesting their potential as
elementary quantum orbitals for constructing larger artificial
quantum systems.

Based on these localized junction orbitals, we further propose a
class of graphene artificial quantum materials constructed from
experimentally realizable three-arm triangulene junctions. The
resulting artificial graphene nanoribbons (aGNRs) exhibit tunable
flat subbands near the Fermi energy arising from CLSs. The
degeneracy of these flat subbands can be controlled by the
artificial ribbon length. In particular, odd-node-width aAGNRs
display strongly direction-dependent transport: flat-subband
states are suppressed when zigzag edges are contacted by
electrodes, whereas they become highly degenerate conducting
channels when armchair edges are contacted. Because the
corresponding CLSs are strongly confined in real space, their
effective Coulomb interactions can become comparable to or larger
than the hopping energies, making these artificial quantum
materials promising platforms for exploring correlation-driven
phenomena, including flat-band magnetism[\onlinecite{LinHH}].

In summary, we establish a unified real-space orbital engineering
framework for understanding and designing localized quantum states
in graphene nanostructures. This approach clarifies the
microscopic origin of interface states in triangulene-based
graphene junctions, explains experimentally observed zero-energy
modes, and provides practical design principles for creating
graphene artificial quantum materials with tunable compact
localized states, flat subbands, and quantum transport
functionalities. These results demonstrate that graphene end
states can serve as programmable quantum orbitals for controlling
electronic transport and designing next-generation artificial
quantum materials.

\textbf{Data availability}\\
The data that supports the finding of this study are available
within the article

\textbf{Conflicts of interest}\\
There are no conflicts to declare.

{\bf Acknowledgments}\\
This work was supported by the National Science and Technology
Council, Taiwan under Contract No. MOST 115-2112-M-008-021.

{}
\mbox{}\\


\appendix

\numberwithin{figure}{section}
\numberwithin{equation}{section}


\section{End states of AGNR Segments}

In Fig.~1, the $n$T-GNR/7-AGNR/$n$T-GNR nanostructures contain
AGNR segments. Therefore, understanding the end states (ESs) of
isolated AGNR segments is essential for interpreting the interface
states of the coupled $n$T-GNR/7-AGNR/$n$T-GNR structures. In
Appendix A, we calculate the energy levels of isolated AGNR
segments as a function of their length along the $y$ direction,
where the length is measured in units of the $R_1$ unit cell, for
various AGNR widths.

As shown in Fig.~A.1(a), the band gap, defined as
$E_g=E_C-E_V=1.323$~eV, reaches a constant value when the length
of the 7-AGNR segment exceeds $y=20~R_1$. This value is consistent
with the intrinsic band gap of an infinite 7-AGNR. Here, $E_C$ and
$E_V$ represent the conduction band minimum and valence band
maximum, respectively. At $y=20~R_1$, two in-gap energy levels
appear at $\Sigma_{0,C}=26.4$~$\mu$eV and
$\Sigma_{0,V}=-26.4$~$\mu$eV. These in-gap states originate from
the ESs localized at the two termini of the AGNR segment. The
overlap between the corresponding wave functions lifts the
degeneracy of the zero-energy modes and results in the finite
energy splitting of $\Sigma_0$.

As the length $y$ decreases, the energy separation between
$\Sigma_{0,C}$ and $\Sigma_{0,V}$ increases significantly. In
particular, we obtain $\Sigma_{0,C(V)}=0.2493(-0.2493)$~eV,
$\Sigma_{0,C(V)}=0.5307(-0.5307)$~eV, and
$\Sigma_{0,C(V)}=1.1183(-1.1183)$~eV for $y=3$, $2$, and $1$,
respectively. These energy levels correspond to the highest
occupied molecular orbital (HOMO) and lowest unoccupied molecular
orbital (LUMO) states of short AGNR segments synthesized
experimentally. The length dependence of $\Sigma_0$ is therefore
important for understanding the interface states $\Sigma_{IF,C}$
and $\Sigma_{IF,V}$ discussed in Figs.~2 and 3.

In Figs.~A.1(b)--A.1(d), the energy levels of 9-AGNR, 13-AGNR, and
15-AGNR segments are calculated. Similar to the 7-AGNR segment,
the 9-AGNR segment exhibits two in-gap energy levels. In contrast,
the 13-AGNR and 15-AGNR segments exhibit four in-gap energy
levels, indicating the emergence of multiple ESs in wider AGNR
segments. The wave functions associated with these degenerate
zero-energy modes exhibit different decay lengths. The states with
longer decay lengths lead to larger energy splittings,
$\Sigma_{C,L}$ and $\Sigma_{V,L}$, whereas the states with shorter
decay lengths exhibit much smaller splittings, $\Sigma_{C,S}$ and
$\Sigma_{V,S}$. This result indicates that the wave functions
corresponding to $\Sigma_{C(V),S}$ are highly localized.

Because fabricating wider AGNR segments remains challenging using
bottom-up synthesis techniques, this work focuses primarily on
$n$T-7-AGNR/7-AGNR/$n$T-7-AGNR nanographene structures reported in
experiments [\onlinecite{DJRizzo},\onlinecite{Pascual}].

\begin{figure}[h]
\centering
\includegraphics[angle=0,scale=0.25]{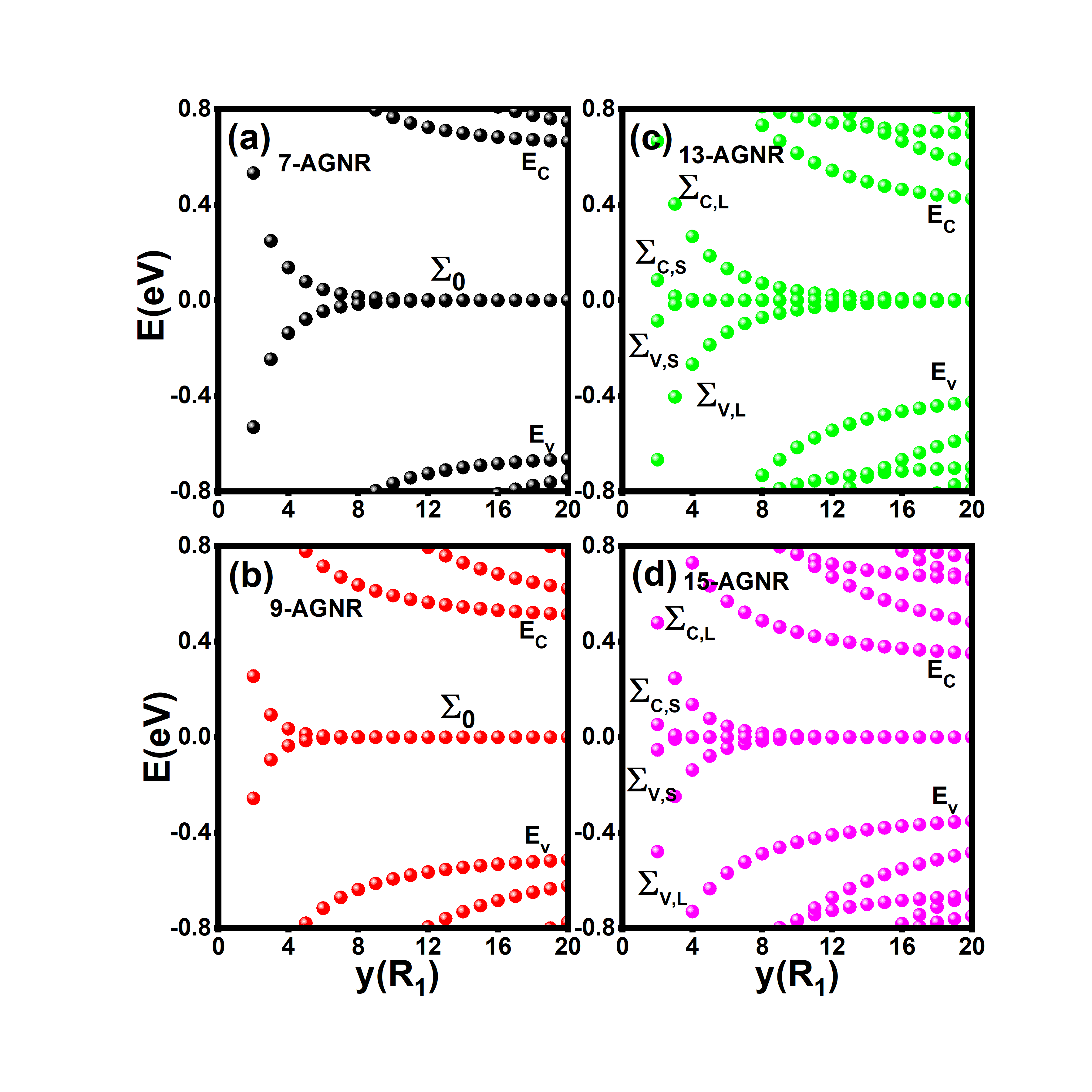}
\caption{Energy levels of AGNR segments as functions of the
segment length, $y$, for different ribbon widths: (a) 7-AGNR, (b)
9-AGNR, (c) 13-AGNR, and (d) 15-AGNR. Here, $y$ denotes the
segment length measured in units of the AGNR unit cell.}
\end{figure}

\section{Physical parameters of nT-GNR/7-AGNR/nT-GNR nanographenes}

The $n$-triangulene/7-AGNR architecture serves as a building block
for artificial quantum materials (AQMs) with honeycomb structures.
The tunable bond length of the 7-AGNR segment between two nodes
containing either mono-orbitals or multiple orbitals acts as an
effective superlattice constant. In Fig.~B.2, we calculate the
electron hopping strength $t_{SL}$, intra-node Coulomb
interactions, and inter-node Coulomb interactions as functions of
the 7-AGNR length $y$, where the length is measured in units of
$3a_{cc}$. The curves with circular and square markers represent
the corresponding physical parameters of the
3T-GNR/$7_y$-AGNR/3T-GNR and 4T-GNR/$7_y$-AGNR/4T-GNR structures,
respectively.

Both $t_{SL}$ curves exhibit exponential decay behavior with
increasing bond length. The decay length of the 4-triangulene
junction is significantly longer than that of the 3-triangulene
junction. This difference can be understood from the charge
density distributions shown in Fig.~5. For the 3T junction, the
charge density of the mono-orbital node state is primarily
localized on the triangulene sites. In contrast, the charge
densities of the multiple orbitals in the 4T junction extend over
the lattice sites of the individual GNR segments (see Fig.~5(i)),
resulting in a longer decay length.

To further clarify the intra- and inter-node Coulomb interactions
($U_0$ and $U_1$), which are evaluated using the expression
\begin{eqnarray}
    &&U_{\ell=0,1}\\ \nonumber
    &=&\frac{1}{4\pi
        \kappa}\sum_{i,j}\frac{|\Psi_{L(R),node}(\textbf{r}_i)|^2
        |\Psi_{L(R),node}(\textbf{r}_j)|^2}{|\textbf{r}_i-\textbf{r}_j|},
\end{eqnarray}
where $\kappa = 4$ is the dielectric constant and $U_{cc} = 4$~eV
is the on-site Coulomb interaction corresponding to double
occupancy in each $p_z$ orbital. The wave functions
$\Psi_{L(R),node}(\textbf{r}_{\ell})$, associated with the left
and right node orbitals, are defined over the regions
$\textbf{r}_{\ell} < \textbf{R}_0$ and $\textbf{r}_{\ell} >
\textbf{R}_0$, respectively. $\textbf{R}_0$ denotes the lattice
sites in the central area of the 7-AGNR segment. The intra-node
electron-electron Coulomb interactions ($U_0$), shown in
Fig.~B.2(b), are relatively insensitive to the variation of the
bond length. This behavior is particularly evident for the
3T-GNR/$7_y$-AGNR/3T-GNR structure and is consistent with the
strongly localized wave functions confined within the triangulene
regions. In contrast, the inter-node Coulomb interactions shown in
Fig.~B.2(c) exhibit a pronounced dependence on the bond length,
reflecting the spatial overlap between localized states located at
different nodes.

\begin{figure}[h]
\centering
\includegraphics[trim=1.cm 0cm 1.cm 0cm,clip,angle=0,scale=0.3]{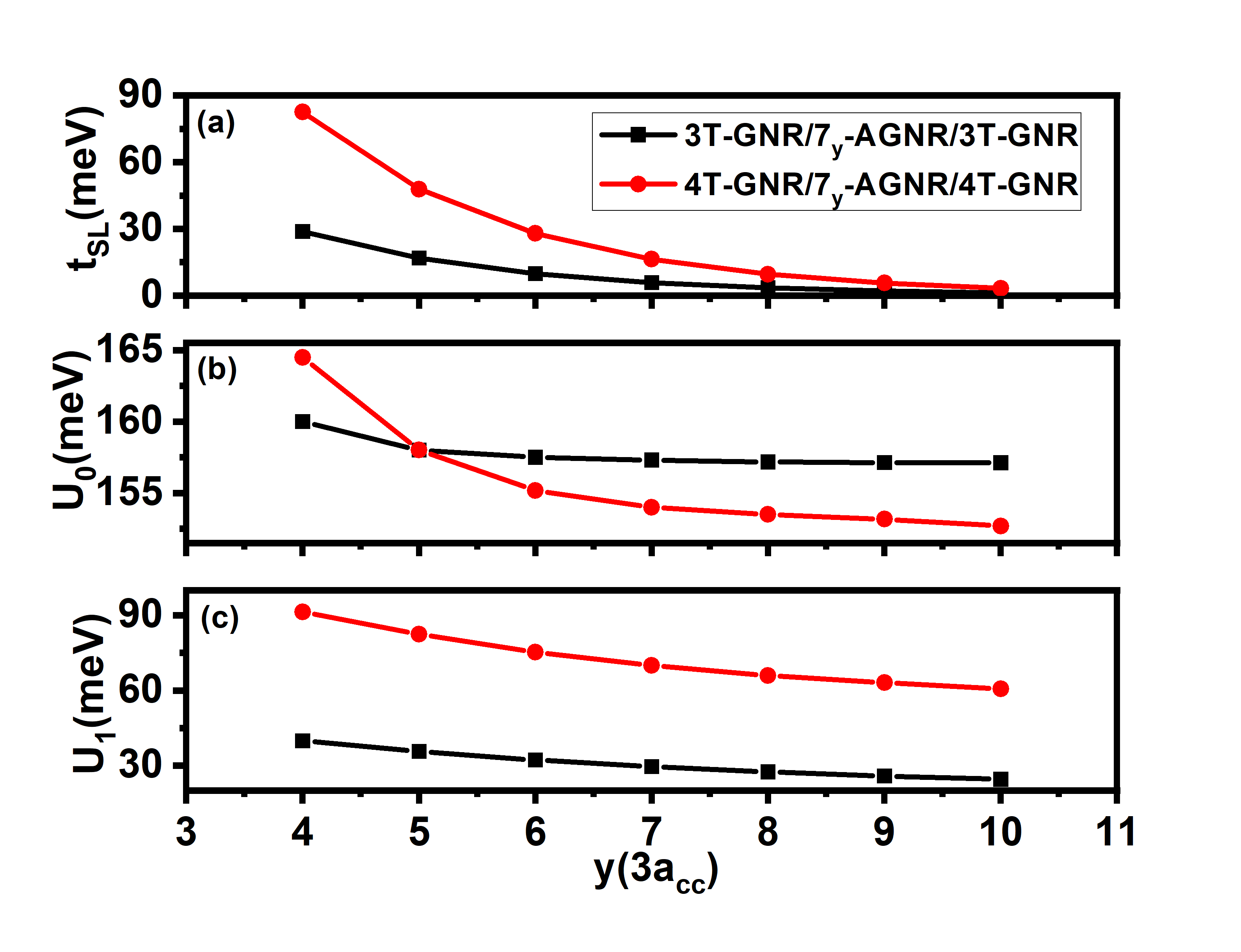}
\caption{(a) Electron hopping strength between two nodes,
$t_{SL}$, (b) intra-node electron Coulomb interaction, $U_0$, and
(c) inter-node electron Coulomb interaction, $U_1$, as functions
of the segment length, $y$. The segment length is expressed in
units of $3a_{cc}$, where $a_{cc}=1.42 \AA$ is the carbon-carbon
bond length.}
\end{figure}

\section{Transmission coefficients of degenerate zero-energy modes in the presence of electron Coulomb interactions}
For completeness, we provide the derivation of the interacting
transmission coefficient used to calculate the STM tunneling
spectra. To investigate the tunneling current in the Kondo regime,
the effect of electron Coulomb interactions on the coupling
between the continuous states of the electrodes and the localized
states of the AAT and BAT molecules must be considered
[\onlinecite{SztenkielD},\onlinecite{BaoZQ}]. In this work, we
focus on the Coulomb blockade regime. Following the
equation-of-motion approach [\onlinecite{DavidK1}], the
transmission coefficient can be written in the closed form:

\begin{equation}
{\cal
T}_{STM}(\epsilon)=\sum_{\ell=1,2}\frac{\Gamma_{\ell,S}*\Gamma_{\ell,D}}{\Gamma_{\ell,S}+\Gamma_{\ell,D}}
(-ImG^{r}_{\ell,\sigma}(\varepsilon)),
\end{equation}

where the single-particle retarded Green's function is given by

\begin{small}
\begin{eqnarray}
G^r_{\ell,\sigma}(\varepsilon)&=&\frac{C_{\ell,1}
}{\epsilon_{\ell}}+\frac{C_{\ell,2}
}{\epsilon_{\ell}-U_{\ell,j}}\\ \nonumber&+&\frac{C_{\ell,3}
}{\epsilon_{\ell}-2U_{\ell,j}}+
\frac{C_{\ell,4}}{\epsilon_{\ell}-U_{\ell}}\\
\nonumber&+&\frac{C_{\ell,5}
}{\epsilon_{\ell}-U_{\ell}-U_{\ell,j}}+
\frac{C_{\ell,6} }{\epsilon_{\ell}-U_{\ell}-2U_{\ell,j}}\\
\nonumber
\end{eqnarray}
\end{small}
Here, $\epsilon_{\ell}=\varepsilon-E_{\ell}+i\Gamma_{\ell}$, and
$\Gamma_{\ell}=(\Gamma_{\ell,S}+\Gamma_{\ell,D})/2$, where
$\Gamma_{\ell,S(D)}$ denotes the tunneling rate between the source
(drain) electrode and the degenerate zero-energy levels of
$\Sigma_0$. The six terms in the single-particle Green's function
correspond to different electronic configurations when an electron
with spin $\sigma$ tunnels through one of the zero-energy modes.
The probability weights for these configurations are

\begin{small}
\begin{eqnarray}
C_{\ell,1}&=&1-N_{\ell,\sigma}-N_{j,\sigma}-N_{j,-\sigma}+ \langle
n_{j,\sigma}n_{\ell,\sigma}\rangle \nonumber \\ &+&\langle
n_{j,-\sigma}n_{\ell,\sigma}\rangle+\langle
n_{j,-\sigma}n_{j,\sigma}\rangle-\langle
n_{j,-\sigma}n_{j,\sigma} n_{\ell,\sigma} \rangle \nonumber \\
C_{\ell,2}&=&N_{j,\sigma}+N_{j,-\sigma}-\langle n_{j,\sigma}
n_{\ell,\sigma}\rangle -\langle n_{j,-\sigma}
n_{\ell,\sigma}\rangle-2\langle n_{j,-\sigma} n_{j,\sigma}\rangle
\nonumber \\ &+&2\langle
n_{j,-\sigma} n_{j,\sigma} n_{\ell,\sigma}\rangle \nonumber \\
C_{\ell,3}&=&\langle n_{j,-\sigma}n_{j,\sigma}\rangle-\langle
n_{j,-\sigma}n_{j,\sigma} n_{\ell,\sigma}\rangle \nonumber\\
C_{\ell,4}&=&N_{\ell,\sigma}- \langle
n_{j,\sigma}n_{\ell,\sigma}\rangle -\langle n_{j,-\sigma}
n_{\ell,\sigma}\rangle \nonumber \\ &+&\langle
n_{j,-\sigma}n_{j,\sigma} n_{\ell,\sigma}\rangle \nonumber \\
C_{\ell,5}&=&\langle n_{j,\sigma} n_{\ell,\sigma}\rangle+\langle
n_{j,-\sigma}n_{\ell,\sigma}\rangle -2\langle
n_{j,-\sigma}n_{j,\sigma} n_{\ell,\sigma}\rangle \nonumber \\
C_{\ell,6}&=&\langle n_{j,-\sigma}n_{j,\sigma}n_{\ell,\sigma}
\rangle \nonumber,
\end{eqnarray}
\end{small}

where $N_{\ell,\sigma}$ represents the single-particle occupation
of energy level $\ell$. The two-particle and three-particle
correlation functions are included self-consistently, satisfying
$\sum_m C_{\ell,m}=1$, which ensures probability conservation. The
transmission formula is valid for temperatures above the Kondo
temperature [\onlinecite{MadhavanV}--\onlinecite{GoldhaberG}].

To calculate the tunneling current in the Coulomb blockade regime,
the single-particle occupation numbers and the two-particle and
three-particle correlation functions must be determined
self-consistently.

\textbf{a: Single-particle occupation number}

The single-particle occupation number
$N_{\ell,\sigma}=N_{\ell,-\sigma}$ is obtained from the
single-particle lesser Green's function:

\begin{eqnarray}
& &N_{\ell,\sigma}\\ \nonumber &=&\int \frac{d\varepsilon}{\pi}~
\frac{\Gamma_{\ell,S}f_S(\varepsilon)+\Gamma_{\ell,D}f_D(\varepsilon)}{\Gamma_{\ell,S}+\Gamma_{\ell,D}}(-ImG^r_{\ell,\sigma}(\varepsilon))
\end{eqnarray}

\textbf{b: Interlevel two-particle correlation functions}

The interlevel two-particle correlation functions include spin
triplet and singlet configurations, $\langle
n_{j,\sigma}n_{\ell,\sigma}\rangle$ and $\langle
n_{j,-\sigma}n_{\ell,\sigma}\rangle$, respectively. Because there
is no electron hopping between the two zero-energy states, these
two correlation functions are equivalent. Therefore, we use the
spin-independent relation $ \langle
n_{j,\sigma}n_{\ell,\sigma}\rangle = \langle
n_{j,-\sigma}n_{\ell,\sigma}\rangle$. The correlation function is
calculated as

\begin{eqnarray}
& & \langle n_{j,\sigma} n_{\ell,\sigma}\rangle\\ \nonumber
&=&\int \frac{d\varepsilon}{\pi}~
\frac{\Gamma_{\ell,S}f_S(\varepsilon)+\Gamma_{\ell,D}f_D(\varepsilon)}{\Gamma_{\ell,S}+\Gamma_{\ell,D}}(-ImG^r_{\ell,j}(\varepsilon)),
\end{eqnarray}

\begin{small}
\begin{eqnarray}
G^r_{\ell,j}(\varepsilon)&=&\frac{F_1}{\epsilon_{\ell}-U_{\ell,j}}+\frac{F_2
}{\epsilon_{\ell}-U_{\ell}-U_{\ell,j}}\\
\nonumber&+&\frac{F_3}{\epsilon_{\ell}-2U_{\ell,j}}+
\frac{F_4}{\epsilon_{\ell}-U_{\ell}-2U_{\ell,j}}\\ \nonumber
\end{eqnarray}
\end{small}
where the probability weights are $F_1=N_{j,-\sigma} -\langle
n_{j,-\sigma}n_{\ell,\sigma}\rangle - \langle
n_{j,-\sigma}n_{j,\sigma}\rangle + \langle
n_{j,-\sigma}n_{j,\sigma}n_{\ell,\sigma}\rangle$, $F_2=\langle
n_{j,-\sigma}n_{\ell,\sigma}\rangle -\langle
n_{j,-\sigma}n_{j,\sigma}n_{\ell,\sigma}\rangle$, $F_3=\langle
n_{j,-\sigma}n_{j,\sigma}\rangle -\langle
n_{j,-\sigma}n_{j,\sigma}n_{\ell,\sigma}\rangle$, and $F_4=\langle
n_{j,-\sigma}n_{j,\sigma}n_{\ell,\sigma}\rangle$.

\textbf{c: Intra-level singlet states}

The two-particle correlation function for intra-level singlet
states is given by:
\begin{small}
\begin{eqnarray}
& &\langle n_{j,-\sigma}n_{j,\sigma}\rangle \nonumber \\
&=&\int
\frac{d\epsilon}{\pi}\frac{\Sigma^{<}_j}{\Gamma_{j,S}+\Gamma_{j,D}}
\Big[-Im( \frac{D_1}{\epsilon_j-U_j}\nonumber
\\&+&\frac{D_2}{\epsilon_j-U_j-U_{\ell,j}}
+\frac{D_3}{\epsilon_j-U_j-2U_{\ell,j}})\Big] \nonumber \\
\end{eqnarray}
\end{small}
In Eq.~(C.6), the lesser self-energy is
$\Sigma^{<}_j=(\Gamma_{j,S}f_S(\varepsilon)+\Gamma_{j,D}f_D(\varepsilon))$.
The corresponding probability weights are defined as:
$D_1=N_{j,\sigma} -\langle n_{\ell,-\sigma}n_{j,\sigma}\rangle -
\langle n_{\ell,\sigma}n_{j,\sigma}\rangle + \langle
n_{\ell,-\sigma}n_{\ell,\sigma}n_{j,\sigma}\rangle$, $D_2=\langle
n_{\ell,\sigma}n_{j,\sigma}\rangle+\langle
n_{\ell,-\sigma}n_{j,\sigma}\rangle-2\langle
n_{\ell,-\sigma}n_{\ell,\sigma}n_{j,\sigma}\rangle$,and
$D_3=\langle n_{\ell,-\sigma}n_{\ell,\sigma}n_{j,\sigma}\rangle$.

\textbf{d:Three Particle Correlation Functions }

The three-particle correlation function is given by:

\begin{small}
\begin{eqnarray}
& &\langle n_{j,-\sigma}n_{j,\sigma}n_{\ell,\sigma}\rangle \nonumber \\
&=&\int
\frac{d\epsilon}{\pi}\frac{\Sigma^{<}_{\ell}}{\Gamma_{\ell,S}+\Gamma_{\ell,D}}
\Big[-Im( \frac{W_1}{\epsilon_{\ell}-2U_{\ell,j}}+
\frac{W_2}{\epsilon_{\ell}-U_{\ell}-2U_{\ell,j}})\Big] \nonumber \\
\end{eqnarray}
\end{small}
The probability weights are defined as: $W_1=\langle
n_{j,-\sigma}n_{j,\sigma}\rangle -\langle
n_{j,-\sigma}n_{j,\sigma}n_{\ell,\sigma}\rangle$, and $W_2=\langle
n_{j,-\sigma}n_{j,\sigma}n_{\ell,\sigma}\rangle$.

Using Eqs.~(C.2)--(C.7), all correlation functions are solved
self-consistently. The obtained solutions are then substituted
into Eq.~(C.1) to calculate the STM tunneling current described by
Eq.~(3). We take $U_{\ell}=0.25$~eV and $U_{\ell,j}=0.08$~eV,
which are calculated from the corresponding wave functions. In
Ref.~[\onlinecite{Pascual}], the applied voltage satisfies $|V_a|
\leq 200$~mV. The probability weights determining the STM
tunneling spectra are mainly governed by the single-particle
occupation numbers and the spin-independent interlevel
two-particle correlation functions.

Note that the transmission coefficient ${\cal
T}_{SD}(\varepsilon)$ in Section II is formulated using
matrix-valued electrode-coupling parameters for the edge-contacted
graphene nanostructure shown in Fig. 1(d). In contrast, ${\cal
T}_{STM}(\varepsilon)$ in Eq. (C.1) is derived from a two-level
Anderson model describing the STM-tip/AAT (or BAT)/metal-substrate
configuration. In this model, the two degenerate zero-energy modes
of the AAT (or BAT) molecule are treated as parallel quantum-dot
levels without an interlevel hopping term, while strong
electron--electron Coulomb interactions are included.

\section{Quantum conductances of aAGNRs with armchair edge sides attached to electrodes}

As discussed in the main text, an additional conductance
contribution originating from the bulk states at $\varepsilon=\pm
t_{SL}$ is observed for 7-aAGNR segments with lengths $N_a=60$
($N_s=15$) and $N_a=68$ ($N_s=17$), as shown in Fig.~14.
Nevertheless, $\sigma_{C(V)}$ does not split into two peaks in the
presence of these bulk states at $\varepsilon=\pm t_{SL}$. This
raises the question of whether the contact properties between the
aAGNR segment and the electrodes can induce hybridization between
the flat subbands and the bulk states.

To clarify this issue, we calculate the electrical conductance of
the 7-aAGNR segment with $N_s=15$ as a function of the chemical
potential for different tunneling rates at zero temperature, as
shown in Fig.~D.1. The electrical conductance $G_e$ associated
with $\sigma_{C(V)}$ remains $16~G_0$ (quantum conductance
$G_0=\frac{2e^2}{h}$) at $\Gamma=1$~meV and $\Gamma=2$~meV when
$\mu = \pm t_{SL}$. For larger tunneling rates, $\Gamma=3$~meV and
$\Gamma=4$~meV, the conductance peaks corresponding to
$\sigma_{C(V)}$ remain higher than $15~G_0$, although their widths
become broader. While the conductance contribution from the bulk
states is significantly enhanced with increasing tunneling rate,
no peak splitting of $\sigma_{C(V)}$ is observed. These results
demonstrate that the $N_s=15$-fold degenerate flat-subband states
remain highly robust CLSs even under strong coupling to the
electrodes.

\begin{figure}[h]
\centering
\includegraphics[trim=1.cm 0cm 1.cm 0cm,clip,angle=0,scale=0.3]{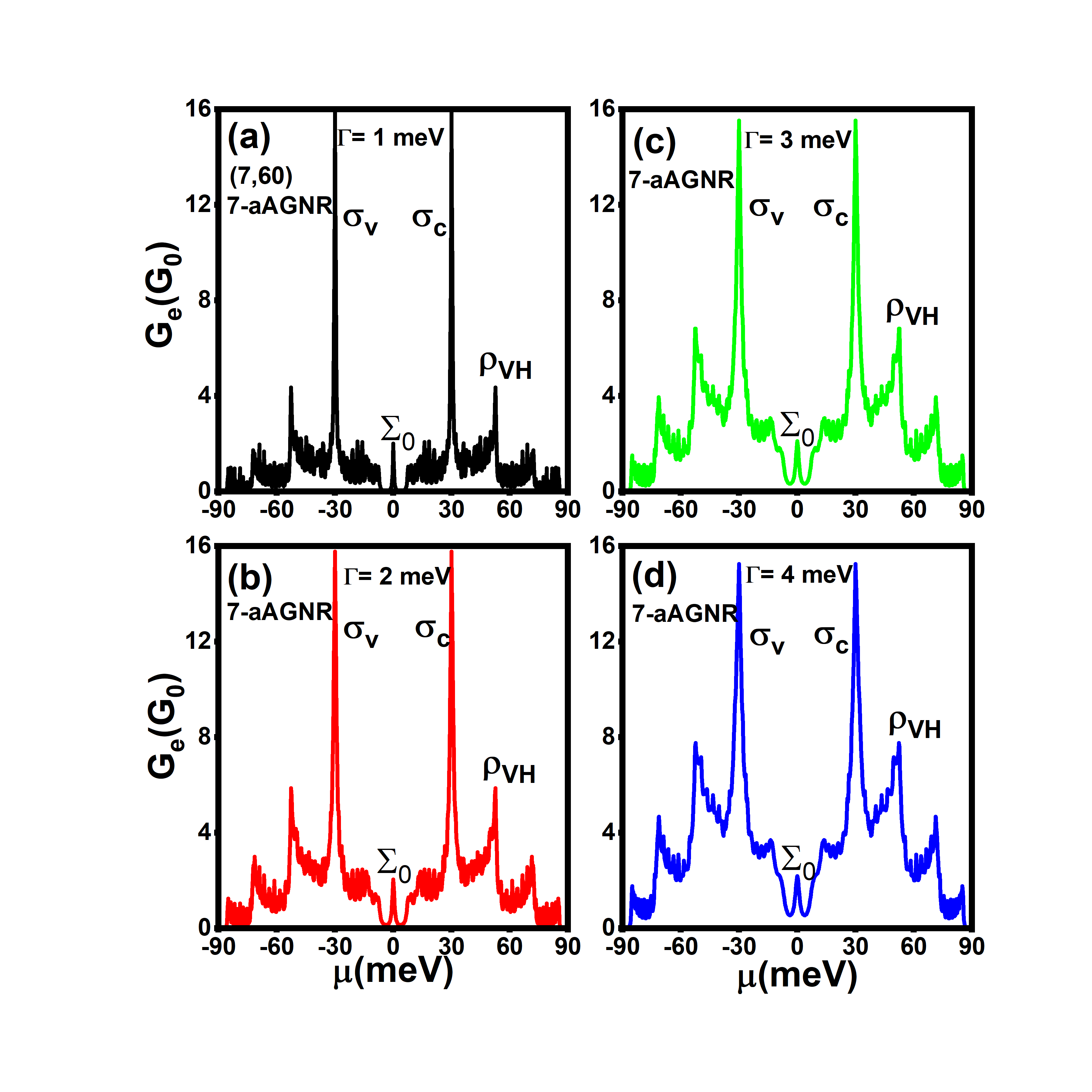}
\caption{Electrical conductance ($G_e$) of 7-aAGNR segments with
length $N_a = 60$, contacted through their armchair edges, as
functions of the chemical potential, $\mu$, at zero temperature
for different tunneling rates: (a) $\Gamma = 1$ meV, (b) $\Gamma =
2$ meV, (c) $\Gamma = 3$ meV, and (d) $\Gamma = 4$ meV.}
\end{figure}

\section{Itinerant-electron-mediated ferromagnetism in flat
bands}

According to Ref.~[\onlinecite{LinHH}], the flat-band
ferromagnetic phase of AGNRs arises from the interplay between
itinerant electrons with $E=\pm t$ and flat bands hosting
localized magnetic moments. It is therefore important to determine
whether bulk states from other subbands can also occur at the
energy levels $E=\pm t_{SL}$ in odd-node aAGNR segments. The
electronic dispersion of an aAGNR segment is given by
\begin{small}
\begin{eqnarray}
& &E(k_{\ell},q_m)\nonumber\\ &=&\pm t_{SL}
\sqrt{1+4\cos^2(q_m)+4\cos(q_m)\cos(k_{\ell}/2)},
\end{eqnarray}
\end{small}
where the transverse and longitudinal wave numbers are,
respectively, $q_m=\frac{m\pi}{N_z+1}$ and
$k_{\ell}=\frac{\ell\pi}{N_s+1}$. Here, $m=1,2,3,\ldots,N_z$ and
$\ell=1,2,3,\ldots,N_s$. The pair $(N_z,N_s)$ denotes an aAGNR
segment with a width of $N_z$ nodes and a length characterized by
$N_s$ unit cells.

When
\begin{equation}
\cos(q_m)+\cos(k_{\ell}/2)=0,
\end{equation}
we obtain $E(k_{\ell},q_m)=\pm t_{SL}$. Consequently, the
condition for bulk states to occur at these energy levels can be
written as
\begin{small}
\begin{equation}
\frac{\ell}{2(N_s+1)}=1-\frac{m}{N_z+1}. \tag{E.2}
\end{equation}
\end{small}
Equation~(E.2) can be rewritten as
$\frac{\ell}{N_s+1}=\frac{m}{(N_z+1)/2}$, with $m<(N_z+1)/2$.

For a 7-aAGNR segment, where $N_z=7$, Eq.~(E.2) shows that an even
value of $N_s$ cannot satisfy this condition. Therefore, at
$E=t_{SL}$, the dispersive subbands do not contribute to the
charge density for 7-aAGNR segments with even $N_s$; the charge
density at this energy originates entirely from the flat subbands,
as shown in Fig.15. In contrast, for $N_s=3,7,11,15,19,\ldots$ and
$N_s=5,9,13,17,\ldots$, the condition in Eq.(E.2) can be
satisfied, and the charge density at $E=t_{SL}$ receives
contributions from both the dispersive bulk states and the flat
subbands, as shown in Fig.~15.

For odd-width aAGNR segments, the itinerant carriers form
standing-wave states along the longitudinal direction. To examine
whether this standing-wave character affects the ferromagnetic
exchange mechanism associated with the flat subbands, we project
the local exchange interaction onto the longitudinal standing-wave
states. The resulting effective exchange Hamiltonian is
\begin{small}
\begin{equation}
H_{ex}= -\sum_{y_i,\ell} J^\ell_{y_i}, \mathbf
S_{F,y_i}\cdot\mathbf s_{it,\ell},
\end{equation}
\end{small}
where $\mathbf S_{F,y_i}$ denotes the localized magnetic moment at
position $y_i$, and $\mathbf s_{it,\ell}$ denotes the spin of the
itinerant electron in the $\ell$th standing-wave state. The
corresponding position-dependent exchange coupling is

\begin{eqnarray}
J^{\ell}_{y_i}& =&\frac{U_0}{N_z+1}|\phi_{\ell}(y_i)|^2\\
\nonumber &=&\frac{2U_0}{(N_s+1)(N_z+1)}
\sin^2\left(\frac{\ell\pi}{N_s+1}y_i\right) \geq 0.
\end{eqnarray}

The standing-wave character therefore introduces a spatial
modulation of the exchange coupling through $|\phi_\ell(y_i)|^2$.
Importantly, this modulation does not change the sign of the
exchange coupling: $J^\ell_{y_i}$ remains nonnegative at every
position $y_i$, although it can vanish at the nodes of the
standing wave. With the convention used in Eq.~(E.3),
$J^\ell_{y_i}>0$ corresponds to ferromagnetic exchange. Thus, the
standing-wave character of the itinerant carriers does not
introduce a sign reversal or frustration of the local exchange
interaction and therefore does not suppress the ferromagnetic
exchange mechanism in odd-width aAGNR segments.







\newpage

\begin{thebibliography}{100}




\bibitem[1]{Novoselovks} K. S. Novoselov, A. K. Geim, S. V. Morozov, D. Jiang, Y. Zhang,
S. V. Dubonos, I. V. Grigorieva, and A. A. Firsov, Electric Field
Effect in Atomically Thin Carbon Films, Science, 2004,
\textbf{306}, 666.

\bibitem[2]{Cai} J. Cai, P. Ruffieux, R. Jaafar, M. Bieri, T. Braun, S. Blankenburg, M. Muoth,
A. P. Seitsonen, M. Saleh, X. Feng, K. Mullen, and Roman Fasel,
Atomically precise bottom-up fabrication of graphene nanoribbons,
Nature, 2010, \textbf{466}, 470 .

\bibitem[3]{ChenYC} Y. C. Chen, T. Cao, C. Chen, Z. Pedramraz, D.
Haberer, D. G. de Oteyza, F. R. Fischer, S. G. Louie and M. F.
Crommie, Molecular bandgap engineering of bottom-up synthesized
graphene nanoribbon heterojunctions, Nat. Nanotechnol., 2015
\textbf{10}, 156.

\bibitem[4]{WangS} S. Y. Wang, L. Talirz, Carlo A. Pignedoli, X. L. Feng, K. Mullen, R.
Fasel and P. Ruffieux, Giant edge state splitting at atomically
precise graphene zigzag edges, Nat. Commun. 2015, \textbf{7},
11507.

\bibitem[5]{LlinasJP} J. P. Llinas,  A. Fairbrother,  Barin G. Borin, W. Shi, K. Lee, S. Wu,
B. Y. Choi, R. Braganza, J. Lear and N. Kau, Short-channel
field-effect transistors with 9-atom and 13-atom wide graphene
nanoribbons, Nat. Commun., 2017, \textbf{8}, 633.

\bibitem[6]{Nestor} N. Merino-Diez, A. Garcia-Lekue, E. Carbonell-Sanroma,`
J. C. Li, M. Corso, L. Colazzo, F. Sedona, D. Sanchez-Portal, J.
I. Pascual, and Dimas G. de Oteyza, Width-Dependent Band Gap in
Armchair Graphene Nanoribbons Reveals Fermi Level Pinning on
Au(111), ACS nano, 2017, \textbf{11}, 11661.

\bibitem[7]{Groning} O. Groning, S. Wang, X. Yao, C.
A. Pignedoli, G. B. Barin, C. Daniels, A. Cupo, V. Meunier, X.
Feng, A. Narita, et al., Engineering of robust topological quantum
phases in graphene nanoribbons, Nature, 2018, \textbf{560}, 209.

\bibitem[8]{Rizzo} D. J. Rizzo, G. Veber, T. Cao, C. Bronner, T. Chen,
F. Zhao, H. Rodriguez, S. G. Louie, M. F. Crommie, and F. R.
Fischer, Topological band engineering of graphene nanoribbons,
Nature, 2018, \textbf{560}, 204.

\bibitem[9]{YanLH} L. H. Yan and P. Liljeroth, Engineered electronic states in
atomically precise artificial lattices and graphene nanoribbons,
Advances in Physics: X, 2019, \textbf{4}, 1651672.


\bibitem[10]{DRizzo} D. J. Rizzo, G. Veber, J. W. Jiang, R. McCurdy, T. Cao
C. Bronner, T. Chen, Steven G. Louie, F. R. Fischer, and M. F.
Crommie, Inducing metallicity in graphene nanoribbons via
zero-mode superlattices, Science,  2020, \textbf{369}, 1597.


\bibitem[11]{DJRizzo} D. J. Rizzo, J. W. Jiang, D. Joshi, G. Veber, C. Bronner, R. A. Durr,
P. H. Jacobse, T. Cao, A. Kalayjian, H. Rodriguez, P. Butler, T.
Chen, Steven G. Louie, F. R. Fischer, and M. F. Crommie,
Rationally designed topological quantum dots in bottom-up graphene
nanoribbons, ACS Nano,  2021, \textbf{15}, 20633.

\bibitem[12]{SunQ} Q. Sun, Y. Yan, X. L. Yao, K. Mullen, A. Narita, R. Fasel, and
P. Ruffieux, Evolution of the topological energy band in graphene
Nanoribbons, J. Phys. Chem. Lett.,  2021, \textbf{12}, 8679.

\bibitem[13]{SongST} S. T Song, Y. Teng, W. C. Tang, Z. Xu, Y. Y. He; J. W. Ruan, T.
Kojima, W. P. Hu, F. J. Giessibl, H. Sakaguchi, S. G. Louie, and
J. Lu, Janus graphene nanoribbons with localized states on a
single zigzag edge, Nature, 2025, \textbf{637}, 580.

\bibitem[14]{LeobandungE} E. Leobandung, L. J. Guo, Y. Wang and S. Y. Chou, Observation of
quantum effects and Coulomb blockade in silicon quantum-dot
transistors at temperatures over 100 K, Appl. Phys. Lett., 1995,
\textbf{67}, 938.


\bibitem[15]{Lloyd} S. Lloyd, Universal Quantum Simulators,
Science, 1996, \textbf{273}, 1073.

\bibitem[16]{Loss} D. Loss and D. P. DiVincenzo, Quantum computation with quantum
dots, Phys. Rev. A, 1998, \textbf{57}, 120.

\bibitem[17]{Ono} K. Ono, D. G. Austing, Y. Tokura and S. Tarucha,
Current Rectification by Pauli Exclusion in a Weakly Coupled
Double Quantum Dot System, Science, 2002, \textbf{297}, 1313.

\bibitem[18]{DiVincenzoDP} D. P. DiVincenzo, Double quantum dot as a quantum
bit, Science, 2005, \textbf{309}, 2173.

\bibitem[19]{WangHM} H. M. Wang, H. S. Wang, C. X. Ma, L. X. Chen, C. X. Jiang, C.
Chen, X. M. Xie, A. P. Li and X. R. Wang, Graphene nanoribbons for
quantum electronics, Nat. Rev. Phys., 2021, \textbf{3}, 791.

\bibitem[20]{PerrinML} J. Zhang, L. Qian, G. B. Barin, P. P. Chen,
K. Mullen, P. Ruffieux, R. Fasel, J. Zhang, M. Calame and M. L.
Perrin, Double quantum dots in atomically-precise graphene
nanoribbons, Mater. Quantum Technol. 2023, \textbf{3}, 036201.


\bibitem[21]{HenriquesJ} J. Henriques, M. Ferri-Cortes and J.
Fernandez-Rossider, Designer Spin Models in Tunable
Two-Dimensional Nanographene Lattices, Nano Lett. 2024,
\textbf{24}, 3355.

\bibitem[22]{HanR} R. Han, J. H. Chen, M. Y. Zhang, J. Gao, Y. C. Xiong, Y. Pan, and T. X. Ma,
Zigzag edge ferromagnetism of triangular-graphene-quantum-dot-like
system, Phys. Rev. B, 2024, \textbf{109}, 075117.

\bibitem[23]{MishraS} S. Mishra, D. Beyer, K. Eimer, R. Ortiz,  J.
Fernandez-Rossider, R. Berger, O. Groning, C. A. Pignedoli, R.
Fasel, X. L. Feng and P. Ruffieux, Collective All-Carbon Magnetism
in Triangulene Dimers, Surface Chemistry, 2020, \textbf{59},
12041.

\bibitem[24]{YuanZG} Z. G. Yuan, X. Y. Zhang, Y. Jiang, X. G.
Qian, Y. Wang, Y. F. Liu, L. Liu, X. Liu, D. Guan, Y. Y. Li, H.
Zheng, C. H. Liu, J. F. Jia, M. P. Qin, P. N. Liu, D. Y. Li, and
S. Y. Wan, Fractional Spinon Quasiparticle in Open-Shell
Triangulene Spin-1/2 Chains, J. Am, Chem. Soc, 2025, \textbf{147},
5004.

\bibitem[25]{VeglianteA} A. Vegliante et al, On-surface Synthesis of a Ferromagnetic Molecular Spin Trimer,
J. Am. Chem. SOC, 2025, \textbf{147}, 19530.


\bibitem[26]{ZhaoCX} C. X. Zhao, L. Yang, J. C. G. Henriues, M.
Ferri-Cortes, G. Catarina, C. A. Pignedoli, J. Ma, X. L. Feng, P.
Rufflieux, J. Fernandez-Rossider and R. Fasel, Spin excitations in
nanographene-based antiferromagnetic spin-1/2 Heisenberg chains,
Nature materials, 2025, \textbf{24}, 722.

\bibitem[27]{YanY} Y. Y. Yan, F. Liu, W. C. Tang, H. X. Wong, B.
Qie, Steven G. Louie and F. R. Fischer, Engineering phase
frustration induced flat bands in an aza-triangulene covalent
Kagome lattice, Nat. Materials, 2026, \textbf{25}, 982.

\bibitem[28]{DavidKMT} David M. T. Kuo, Topological interface states and nonlinear
thermoelectric performance in armchair graphene nanoribbon
heterostructures, RSC Adv., 2026, \textbf{16}, 4680.

\bibitem[29]{TamakiG} G. Tamaki, T. Kawakami and M. Koshino,
Topological junction states and their crsytalline network in
chiral symmetric systems: application to graphene nanoribbons,
Phys. Rev. B, 2020, \textbf{101}, 205311.

\bibitem[30]{GuYW} Y. W. Gu, Z. Qiu and K. Mullen, Nanographenes and Graphene
Nanoribbons as Multitalents of Present and Future Materials Science,
J. Am. Chem. Soc , 2022, \textbf{144}, 11499.

\bibitem[31]{LiebEH} E. H. Lieb, Two Theorems on the Hubbard Model, Phys. Rev. Lett. 1989, \textbf{62}, 1201.

\bibitem[32]{Rossier} J. Fermandez-Rossier and J. J. Palacios,
Magnetism in Graphene Nanoislands, Phys. Rev. Lett. 2007,
\textbf{99}, 177204.

\bibitem[33]{Yazyev} O. V. Yazyev, Emergence of magnetism in
graphene materials and nanostructures, Rep. Prog. Phys. 2010,
\textbf{73}, 056501.

\bibitem[34]{OrtizR} R. Ortiz, G. Catarina and J.
Fernandez-Rossider, Theory of triangulene two-dimensional
crystals, 2D Materials, 2023, \textbf{10}, 015015.

\bibitem[35]{MadailL} L. Madail, R. G. Dias and J.
Fernandez-Rossider, Exotic edge states of $C_3$ high-fold fermions
in honeycomb lattices, Phys. Rev. Research, 2024, \textbf{6},
043262.

\bibitem[36]{Pascual} F. Romero-Lara, M. Vilas-Varela, R. Ortiz,
M. Kumar, A. Vegliante, L. Gomez-Rodrigo, J. P. Calupitan, D.
Soler, N. Friedrich, D. F. Wang, J. Ortuzar, S. Trivini, F.
Schulz, T. Frederiksen, P. Jelinek, D. Pena, and J. I. Pascual,
Topological Engineering of a Frustrated Antiferromagnetic
Triradical in Aza-Triangulene Architecture, arXiv:2512.10869v1.


\bibitem[37]{DelplaceP} P. Delplace, D. Ullmo and G. Montambaux, The Zak phase and the
existence of edge states in graphene, Phys. Rev. B 2011, 84,
195452.

\bibitem[38]{CaoT} T. Cao, F. Z. Zhao, and S. G. Louie, Topological Phases in Graphene Nanoribbons:
Junction States, Spin Centers and Quantum Spin Chains, Phys. Rev.
Lett. 2017, \textbf{119}. 076401.

\bibitem[39]{LinKS} K. S. Lin and M. Y. Chou, Topological
Properties of Gapped Graphene Nanoribbons with Spatial Symmetries,
Nano Lett, 2018, \textbf{ 18}, 7254.


\bibitem[40]{Kariyado} T. Kariyado and  X.  Hu, Topological States Characterized
by Mirror Winding Numbers in Graphene with Bond Modulation,
Scientific Report, 2017, \textbf{7}, 16515.


\bibitem[41]{Bardarsonjh} J. W. Rhim, J. H. Bardarson and R. J. Slager,
Unified bulk-boundary correspondence for band insulators, Phys.
Rev. B  2018, \textbf{97}, 115143.

\bibitem[42]{JiangJW} J. W. Jiang and Steven G. Louie, Topology Classification using Chiral
Symmetry and Spin Correlations in Graphene Nanoribbons, Nano Lett,
2021, \textbf{21}, 197.


\bibitem[43]{MorishitaN} N. Morishita, K. Komatsu, M. Kitatani, and
K. Kusakabe, Modulated Dirac bands and integer hopping ratios in a
honeycomb lattice of phenalenyl-tessellation molecules, Phys Rev.
B, 2025, \textbf{111}, 235117.

\bibitem[44]{LuCH} C. H. Lu, and E. Y. T Li, A new
graph theory to unravel the bulk-boundary correspondence of
graphene nanoribbons, Carbon, 2024, \textbf{230}, 119624.

\bibitem[45]{ChenBH} B. H. Chen and D. W. Chiou, An elementary rigorous proof of
bulk-boundary correspondence in the generalized
Su-Schrieffer-Heeger model, Phys. Lett. A, 2020, \textbf{380},
126168.

\bibitem[46]{Maalysheva} L. Maalysheva and A. Onipko, Spectra of $\pi$ Electrons in Graphene as a Macromolecule,
Phys. Rev. Lett. 2008, \textbf{100}, 186806.

\bibitem[47]{OnipkoA} A. Onipko, Spectra of $\pi$ electrons in
graphene as an aletrnant macromoleculae and its specific features
in quantum conductance, Phys. Rev. B  2008, \textbf{78}, 245412.


\bibitem[48]{NikolaevAV} A. V. Nikolaev, A. V. Bibikov, A. V. Avdeenkov, I. V. Bodrenko, and E. V. Tkalya,
Electronic and transport properties of rectangular graphene
macromolecules and zigzag carbon nanotubes of finite length, Phys.
Rev. B, 2009, \textbf{79}, 045418.

\bibitem[49]{RozhkovAV} A. V. Rozhkov and F. Nori, Exact wave functions for an electron on a graphene triangular quantum dot,
Phys. Rev. B, 2010, \textbf{81}, 155401.

\bibitem[50]{PotaszP} P. Potasz, A. D. Guclu and P. Hawrylak, Zero-energy states in triangular and trapezoidal graphene
structures, Phys. Rev. B, 2010, \textbf{81}, 033403.

\bibitem[51]{TalkachovA} A. Talkachov and E. Babaev, Wave functions and edge states in rectangular honeycomb lattices revisited: Nanoflakes,
armchair and zigzag nanoribbons, and nanotubes, Phys. Rev. B,
2023, \textbf{107}, 045419.

\bibitem[52]{LopezSancho} M. Pilar Lopez-Sancho and M. Carmen
Munoz, Topologically protected edge and confined states in finite
armchair graphene nanoribbons and their junctions, Phys. Rev. B,
2021, \textbf{104}, 245402.

\bibitem[53]{MariO} M. Ohfuchi, and S. Sato, Energetics and magnetism of topological graphene
nanoribbons, J. Appl. Phys., 2021, \textbf{129}, 064305.

\bibitem[54]{ZhaoFZ} F. Z. Zhao, T. Cao, and Steven G. Louie, Topological Phases in Graphene Nanoribbons Tuned by Electric Fields
Phys. Rev. Lett. 2021, \textbf{127}, 166401.

\bibitem[55]{JacobsePH} M. J. J. Mangnus, F. R. Fischer, M. F. Crommie, I. Swart and P. H.
Jacobse, Charge transport in topological graphene nanoribbons and
nanoribbon heterostructures, Phys. Rev. B, 2022, \textbf{105},
115424.

\bibitem[56]{Traverso} S. Traverso, M. Sassetti, and N. T. Ziani, Role of the edges in a quasicrystalline Haldane model,
Phys. Rev. B, 2022, \textbf{106}, 125428.

\bibitem[57]{TepliakovNV} N. V. Tepliakov, J. Lischner, E. Kaxiras, A. A. Mostofi, and M. Pizzochero,
Unveiling and Manipulating Hidden Symmetries in Graphene
Nanoribbons, Phys. Rev. Lett., 2023, \textbf{130}, 026401.

\bibitem[58]{HuangA} A. H. Huang, S. S. Ke, J. H. Guan, J. Li,
and W. K. Lou, Strain-induced topological phase transition in
graphene nanoribbons, Phys. Rev.  B, 2024, \textbf{109}, 045408.

\bibitem[59]{DavidKuo} David M. T. Kuo, Topological states in finite graphene nanoribbons tuned by electric
fields, J. Phys: Condens. Matter, 2025, \textbf{37}, 085304.

\bibitem[60]{Ostmeyer} J. Ostmeyer, L. Razmadze, E. Berkowitz, T.
Luu and U. G. Meibner, Effective theory for graphene nanoribbons
with junctions, Phys. Rev. B, 2024, 109, 195135.

\bibitem[61]{Abdelsalam} H. Abdelsalam, D. Corona, R. B. Payod, M. A.
S. Sakr, Omar H. Abd-Elkader, Q. F Zhang, and V. A. Saroka,
Topological Junction States in Graphene Nanoribbons: A Route to
Topological Chemistry, Nano Lett. 2025, \textbf{25}, 10594.

\bibitem[62]{ZdetsisAD} A. D. Zdetsis, Peculiar electronic
properties of wider armchair graphene nanoribbons:Multiple
topological end states and "phase transistion", Carbon, 2023,
\textbf{210}, 118042.

\bibitem[63]{Kuo3} David M. T. Kuo, Charge transport through the multiple end zigzag
edge states of armchair graphene nanoribbons and heterojunctions,
RSC Adv., 2024, \textbf{14}, 20113.

\bibitem[64]{DavidMTK1} David M. T. Kuo, Room-Temperature Pauli Spin Blockade and Current
Rectification in 15-13-15 Armchair Graphene Nanoribbon Heterostructures,
Nanoscale, 2025, \textbf{17}, 18920.

\bibitem[65]{SanzSO} S. Sanz and D. Sanchez-Portal, Predicting interface
and spin states in armchair graphene nanoribbon junctions, Phys,
Rev. B, 2026, 113, 235434.


\bibitem[66]{Danieli} C. Danieli, A. Andreanov, and S. Flach,
Many-body flatband localization, Phys. Rev. B, 2020, \textbf{102},
041116 (R).

\bibitem[67]{DanieliC} C. Danieli, A. Andreanov, and S. Flach,
Many-body localization transition from flat-band fine tuning,
Phys. Rev. B, 2022, \textbf{105}, L041113.

\bibitem[68]{RomerRA} C. Danieli, J. Liu, R. A. Romer and R. A.
Vicencio, Quantum storage with flat band, Phys. Rev. Lett, 2026,
\textbf{136}, 066302.


\bibitem[69]{ChangYC} Y. C. Chang, J. N. Schulman, G. Bastard, and Y. Guldner, Effects of
quasi-interface states in HgTe-CdTe superlattices. Phys. Rev. B,
1985, \textbf{31}, 2557.

\bibitem[70]{ZhangSC} B. Andrei Bernevig, T. L. Hughes and S. C.
Zhang, Quantum spin hall effect and topological phase transition
in HgTe quantum wells, Science, 2006, 314, 1757.

\bibitem[71]{HasanMZ} M. Z. Hasan and C. L. Kane,
Colloquium:Topological insulators, Rev. Mod. Phys. 2010,
\textbf{82}, 3045.

\bibitem[72]{AnisimovVI} V. I. Anisimov, A. I. Poteryaev, M. A.
Korotin, A. O. Anoknin and G. Kotliar, First principle
calculations of the electronic structure and spectra of strongly
corelated systems: dynamical mean-field theory, J. Phys: Condens.
Matter 1997, \textbf{9}, 7359.

\bibitem[73]{NakadaK} K. Nakada, M. Fujita, G. Dresselhaus and M. S. Dresselhaus, Edge
state in graphene ribbons: Nanometer size effect and edge shape
dependence, Phys. Rev. B, 1996, \textbf{54}, 17954.

\bibitem[74]{WakabayashiK} K. Wakabayashi, M. Fujita, H. Ajiki, and M. Sigrist, Electronic
and magnetic properties of nanographite ribbons, Phys. Rev. B,
1999, \textbf{59}, 8271.

\bibitem[75]{WakabayashiK2} K. Wakabayashi, K Sasaki, T. Nakanishi and T. Enoki, Electronic
states of graphene nanoribbons and analytical solutions, Sci.
Technol. Adv. Mater., 2010, \textbf{11}, 054504.


\bibitem[76]{AllenMT} M. T. Allen, J. Martin, and A. Yacoby, Gate-defined quantum
confinement in suspended bilayer graphene, Nat. commun., 2012,
\textbf{3}, 934.

\bibitem[77]{DasSarmaS} S. Das Sarma, S. Adam, E. H. Hwang, and E. Rossi, Electronic
transport in two-dimensional graphene, Rev. Mod. Phys., 2011,
\textbf{83}, 407.

\bibitem[78]{KotovVN} V. N. Kotov, B. Uchoa, V. M. Pereira, F. Guinea, and A. H.
Castro Neto, Electron-Electron Interactions in Graphene: Current
Status and Perspectives, Rev. Mod. Phys., 2012, \textbf{84}, 1067.


\bibitem[79]{VolkC} C. Volk, C. Neumann, S. Kazarski, S. Fringes, S. Engels, F. Haupt,
A. Mueller, and C. Stampfer, Probing relaxation times in graphene
quantum dots, Nat. commun., 2013, \textbf{4}, 1753.

\bibitem[80]{GerardotBD} M. Brotons-Gisbert, B. Artur, S. Kumar, R. Picard, R. Proux, M.
Gray, K. S. Burch, K. Watanabe, T. Taniguchi, and B. D. Gerardot,
Coulomb blockade in an atomically thin quantum dot coupled to a
tunable Fermi reservoir, Nat. nanotechonol., 2019, \textbf{14},
442.

\bibitem[81]{SunQF} Q. F. Sun and X. C. Xie, CT-Invariant Quantum Spin Hall Effect in
Ferromagnetic Graphene, Phys. Rev. Lett., 2010, \textbf{104},
066805.


\bibitem[82]{Kuo5} David M. T. Kuo and Y. C. Chang, Contact Effects on
Thermoelectric Properties of Textured Graphene Nanoribbons,
Nanomaterials, 2022, \textbf{12}, 3357.


\bibitem[83]{SolsF} F. Sols, F. Guinea, and A. H. Castro Neto,
Coulomb Blockade in Graphene Nanoribbons, Phys. Rev. Lett., 2007,
\textbf{99}, 166803.

\bibitem[84]{ToddK} K. Todd, H. T. Chou, S. Amasha and D.
Goldhaber-Gordon, Quantum Dot behavior in Graphene
Nanoconstrictions, Nano Lett., 2009, \textbf{9}, 416.


\bibitem[85]{Cernevics} K. Cernevics, O. V. Yazyev, and M.
Pizzochero, Electronic transport across quantum dots in graphene
nanoribbons: Toward built-in gap-tunable metal-semiconductor-metal
heterojunctions, Phys. Rev. B. 2020, \textbf{102}, 201406(R).


\bibitem[86]{Abbassi} M. E. Abbassi, M. L. Perrin, G. B. Barin, S. Sangtarash, J. Overbeck,
O. Braun, C. J. Lambert, Q. Sun, T. Prechtl, A. Narita, K. Mullen,
P. Ruffieux, H. Sadeghi, R. Fasel, and M. Calame, Controlled
Quantum Dot Formation in Atomically Engineered Graphene Nanoribbon
Field-Effect Transistors, ACS Nano, 2020, \textbf{14}, 5754.

\bibitem[87]{Huangwh} W. H. Huang, O. Braun, David I. Indolese, G.
B. Barin, G. Gandus, M. Stiefel, A. Olziersky, K. Mullen, M.
Luisier, D. Passerone, P. Ruffieux, C. Schonenberger, K. Watanabe,
T. Taniguchi, R. Fasel, J. Zhang, Michel Calame, and M. L. Perrin,
Edge Contacts to Atomically Precise Graphene Nanoribbons, ACS
Nano, 2023, \textbf{17}, 18706.

\bibitem[88]{ZhangJain} J. Zhang, L. Qian, G. B. Barin, A. H. S. Daaoub,
P. P. Chen, K. Mullen, S. Sangtarash, Pascal Ruffieux, R. Fasel,
Hatef Sadeghi, J. Zhang, M. Calame and M. L. Perrin, Contacting
individual graphene nanoribbons using carbon nanotube electrodes,
Nat. electron., 2023, \textbf{6}, 572.

\bibitem[89]{DavidK1} David M. T. Kuo, S. Y. Shiau, and Y. C.
Chang, Theory of spin blockade, charge ratchet effect, and
thermoelectrical behavior in serially coupled quantum dot system,
Phys. Rev. B, 2011, \textbf{84}, 245303.

\bibitem[90]{DavidK2} C. C. Chen, Y. C. Chang and David M. T. Kuo,
Quantum interference and electron correlation in charge transport
through triangular quantum dot molecules, Phys. Chem. Chem. Phys.,
2015, \textbf{17}, 6606.

\bibitem[91]{DavidK3} C. C. Chen, David M. T. Kuo and Y. C. Chang, Quantum interference and structure-dependent
orbital-filling effects on the thermoelectric properties of
quantum dot molecules, Phys. Chem. Chem. Phys., 2015, \textbf{17},
19386.

\bibitem[92]{Kuo6} David M. T. Kuo, Temperature-stable tunneling
current in serial double quantum dots: insights from
nonequilibrium Green's functions and Pauli spin blockade, Phys.
Chem. Chem. Phys., 2025, \textbf{27}, 5238.

\bibitem[93]{MadhavanV} V. Madhavan, W. Chen, T. Jamneala, M. F. Crommie, N. S.
Wingreen, Tunneling into a Single Magnetic Atom: Spectroscopic
Evidence of the Kondo Resonance, Science, 1998, \textbf{280}, 567.

\bibitem[94]{Goldhaber} D. Goldhaber-Gordon, H.
Shtrikman, D Mahalu, D. Abusch-Magder, U. Meirav, M. A. Kastner,
Kondo effect in a single-electron transistor, Nature, 1998,
\textbf{391}, 156.

\bibitem[95]{GoldhaberG}D. Goldhaber-Gordon, J. Gores, M. A. Kastner, Hadas
Shtrikman, D. Mahalu, and U. Meirav, From the Kondo Regime to the
Mixed-Valence Regime in a Single-Electron Transistor, Phys. Rev.
Lett., 1998, \textbf{81}, 5225.

\bibitem[96]{SztenkielD} D. Sztenkiel and R. Swirkowicz, Conductance of a double quantum dot system
in the Kondo regime in the presence of inter-dot coupling and
channel mixing effects, J. Phys.: Condens. Matter 2007,
\textbf{19}, 256205.

\bibitem[97]{BaoZQ} Z. Q. Bao, A. M. Guo and Q. F. Sun, Orbital Kondo effect in a parallel double quantum
dot, J. Phys.: Condens. Matter 2014, \textbf{26}, 435301.


\bibitem[98]{NaritsukaM} M. Naritsuka, T. Machida, S. Asano, Y.
Yanase and T. Hanaguri, Superconductivity controlled by twist
angle in monolayer NbSe2 on graphene, Nature Physics, 2025,
\textbf{21}, 746.

\bibitem[99]{SpataruCD} C. D. Spataru, W. Pan, and A. Cerjan, Topological
Phenomena in Artificial Quantum Materials Revealed by Local Chern
Markers, Phys. Rev. Lett. 2025, \textbf{134}, 126601.

\bibitem[100]{ShenLG} L. G. Shen
, M. Lin, C. Y. Y. Lin, D. Xiao and T. Cao, Realization of
fermionic Laughlin state on a quantum processor, Nature
Communications, 2026, \textbf{17}, 4919.


\bibitem[101]{JoGB} G. B. Jo, J. Guzman, C. K. Thomas et al, Ultracold Atoms in a Tunable Optical Kagome Lattice,
Phys. Rev. Lett. 2012, \textbf{108}, 045305.

\bibitem[102]{TaieS} S. Taie, H. Ozawa, T. Ichinose, T. Nishio, S. Nakajima and Y. Takahashi,
Coherent driving and freezing of bosonic matter wave in an optical
Lieb lattice, Science Advances, 2015, \textbf{1}, 1500854.

\bibitem[103]{RodrigoA} R. A. Vicencio, C. Cantillano, L. Morales-Inostroza, Bastian Real, S. Weimann,
A. Szameit, and M. I. Molina, Observation of Localized States in
Lieb Photonic Lattices, Phys. Rev. Lett., 2015, \textbf{114},
245503.

\bibitem[104]{Mukherjee}S. Mukherjee, A. Spracklen, D. Choudhury,
N. Goldman, P. Ohberg, E. Andersson, and R. R. Thomson,
Observation of a Localized Flat-Band State in a Photonic Lieb
Lattice, Phys. Rev. Lett., 2015, \textbf{114}, 245504.

\bibitem[105]{LeykamD} D. Leykam, A. Andreanov, and S. Flach,
Artificial flat band systems: from lattice models to experiments,
Advances in Physics: X, 2018, \textbf{3}, 1473052.

\bibitem[106]{EzziM} M. M. Al Ezzi, J. X. Hu, A. Ariando, F.
Guinea, and S. Adam, Topological Flat Bands in Graphene
Super-Moire Lattices, Phys. Rev. Lett. 2024, \textbf{132}, 126401.


\bibitem[107]{YinRT} R. T. Yin, X. Y. Meng, X. J. Zhao, J. N. Wang,
X. Q. Wang, Q. Chen, J. Meng, Z. Y. Wang, Y. F. Liang, Y. Z Tan,
B. Li, W. Hu, Q. X. Li, S. Tan, C. Ma, J. L. Yang, and B. Wang,
Tailoring Near Fermi-Level Topological Flatbands in Clar's Goblet
Graphene Nanoribbons through Regioselective Cyclization of
Five-Membered Rings, J. Am. Chem. Soc. 2025, \textbf{147}, 34303.


\bibitem[108]{SevincliH} H. Sevincli, M. Topsakal, and S. Ciraci, Superlattice structures
of graphene-based armchair nanoribbons, Phys. Rev. B 2008,
\textbf{78}, 245402.

\bibitem[109]{ZhengHX} H. X. Zheng, Z. F. Wang, T. Luo, Q. W. Shi, and
J. Chen, Analytical study of electronic structure in armchair
graphene nanoribbons, Phys. Rev. B 2007, \textbf{75}, 165414.

\bibitem[110]{LinHH} H. H. Lin, T. Hikihara, H. T. Jeng, B. L, Huang, C. Y. Mou and
X. Hu, Ferromagnetism in armchair graphene nanoribbons, Phys. Rev.
B 2009, \textbf{79}, 035405.

\bibitem[111]{AlmeidaPA} P. A. Almeida, L. S. Sousa, T. M. Schmidt
and G. B. Martins, Ferromagnetism in armchair graphene nanoribbon
heterostructures, Phys. Rev. B 2022, \textbf{105,} 054416.

\bibitem[112]{HodO} Oded Hod, Juan E. Peralta,and Gustavo E. Scuseria, Edge effects in
finite elongated graphene nanoribbons, Phys Rev B 2007,\textbf{
76}, 233401.

\bibitem[113]{SuiM} M. Sui, G. Chen, L. Ma, W. Y. Shan, D. Tian,
K. Watanabe, T. Taniguchi, X. F. Jin, W. Yao, D. Xiao and Y. B.
Zhang, Gate-Tunable Topological Valley Transport in Bilayer
Graphene. Nat. Phys. 2015, \textbf{11}, 1027.


\bibitem[114]{WangJN} J. N. Wang, W. W. Chen, Z. Y. Wang, J. Meng,
R. T. Yin, M. G. Chen, S. J. Tan, C. X, Ma, Q. X. Li and  B. Wang,
Designer topological flat bands in one-dimensional armchair
graphene antidot lattices, Phys. Rev. B 2024, \textbf{110},
115138.

\bibitem[115]{YuanXY} X. Y. Yuan, J. W. Xin and S. Zhu,
Topological flat-band engeering in frustarted graphene anti-dot
lattices, Phys. Rev. B, 2025, \textbf{112}, 235435.


\bibitem[116]{Bergman} D. L. Bergman, C. J. Wu, and L. Balents,
Band touching from real-space topology in frustrated hopping
models, Phys. Rev. B, 2008, \textbf{78}, 125104.

\bibitem[117]{RhimJW} J. W. Rhim, and B. J. Yang,
Classification of flat bands according to the band-crossing
singularity of Bloch wave functions, Phys. Rev. B, 2019,
\textbf{99}, 045107.

\bibitem[118]{Morfonios} C. V. Morfonios, M. Rontgen, M. Pyzh, and P. Schmelcher,
Flat bands by latent symmetry, Phys. Rev. B, 2021, \textbf{ 104},
035105.

\bibitem[119]{Guedon} C. M. Guedon,  H. Valkenier, T. Markussen,
K. S. Thygesen, J. C. Hummelen, and Sense Jan van der Molen,
Observation of quantum interference in molecular charge transport,
Nature nanotechnology 2012,\textbf{7}, 305.

\bibitem[120]{Markussen} T. Markussen, R. Stadler and K. S. Thygesen, The relation between structure and quantum
interfernce in a single molecule junctions, Nano. Lett. 2010,
\textbf{10}, 4260.

\bibitem[121]{Cardamone} D. M. Cardamone, C. A. Stafford and S.
Mazumdar, Controlling quantum transport through a single molecule,
Nano. Lett. 2006, \textbf{6}, 2422.


\end{thebibliography}
\end{document}